\documentclass[final,referee]{paper}

\usepackage{todonotes}
\usepackage[mathlines]{lineno}
\newcommand*\patchAmsMathEnvironmentForLineno[1]{%
  \expandafter\let\csname old#1\expandafter\endcsname\csname #1\endcsname
  \expandafter\let\csname oldend#1\expandafter\endcsname\csname end#1\endcsname
  \renewenvironment{#1}%
     {\linenomath\csname old#1\endcsname}%
     {\csname oldend#1\endcsname\endlinenomath}}%
\newcommand*\patchBothAmsMathEnvironmentsForLineno[1]{%
  \patchAmsMathEnvironmentForLineno{#1}%
  \patchAmsMathEnvironmentForLineno{#1*}}%
\AtBeginDocument{%
\patchBothAmsMathEnvironmentsForLineno{equation}%
\patchBothAmsMathEnvironmentsForLineno{align}%
\patchBothAmsMathEnvironmentsForLineno{flalign}%
\patchBothAmsMathEnvironmentsForLineno{alignat}%
\patchBothAmsMathEnvironmentsForLineno{gather}%
\patchBothAmsMathEnvironmentsForLineno{multline}%
}
\usepackage{empheq}
\usepackage{amsmath}
\usepackage{mathtools}
\usepackage{amssymb} %
\usepackage{mathrsfs} %
\usepackage{IEEEtrantools} %
\usepackage{fp}   %

\usepackage[subnum]{cases} %
\usepackage{cancel}

\usepackage{xstring} %

\def\u2#1{\underline{\underline{#1}}} %
\newcommand{\bs}[1]{\boldsymbol{#1}} %

\newcommand{\fpdv}[1]{\partial_{#1}}

\usepackage[shortlabels]{enumitem} %
\usepackage{pifont}     %

\usepackage{parskip} %
\usepackage[normalem]{ulem} %
\usepackage{accents}
\usepackage{csquotes} %
\def\colorize<#1>{\temporal<#1>%
{\color{black!30}}%
{\color{red}}%
{\color{black}}%
}
\usepackage{varwidth,lipsum} %
\usepackage{tabto} %
\usepackage{tabstackengine} %

\usepackage{fullpage}

\usepackage[percent]{overpic} %
\usepackage[rightcaption]{sidecap}
\usepackage{caption}
\usepackage{subfig} %
\usepackage{graphicx} %

\usepackage{epstopdf} %
\usepackage{latexsym}%
\usepackage{keyval}%
\usepackage{ifthen}%
\usepackage{moreverb}%
\usepackage[subfolder]{gnuplottex}%

\usepackage{tikz}
\usetikzlibrary{decorations.markings}

\DeclareRobustCommand{\fillcircle}[1]{
\begin{tikzpicture}[baseline=-0.5ex]
  \draw[#1 ,fill= #1 ] (0,0) circle (.5ex);
\end{tikzpicture}
}
\DeclareRobustCommand{\lineplain}[1]{
\begin{tikzpicture}[baseline=-0.5ex]
  \draw[line width=0.5mm, Redone] (0,0) -- (0.5,0);
\end{tikzpicture}
}
\DeclareRobustCommand{\linedot}[1]{
\begin{tikzpicture}[baseline=-0.5ex]
  \draw[line width=0.5mm, dash pattern=on 2pt off 1pt, Blueone] (0,0) -- (0.5,0);
\end{tikzpicture}
}
\graphicspath{{./}}

\usepackage[
style=numeric-comp,
citestyle=numeric,
backend=biber,
firstinits=true,%
useprefix=true,%
sorting=none, %
sortcites=true,
bibencoding=utf-8,
maxnames=15,
minnames=1,
maxbibnames=15,%
minbibnames=14,
maxcitenames=2,%
mincitenames=2,
hyperref=auto,
backref=false,
backrefstyle=all+,
isbn=false,
eprint=false,
url=false,
doi=false]{biblatex}

\renewbibmacro*{volume+number+eid}{%
  \printfield{volume}%
  \printfield{number}%
  \printfield{eid}}

\DeclareNameAlias{sortname}{last-first}
\DeclareNameAlias{default}{last-first}

\renewbibmacro{in:}{}

\DeclareFieldFormat[article]{volume}{\mkbibbold{#1}}%
\DeclareFieldFormat[article]{number}{#1}%
\DeclareFieldFormat[article]{title}{#1} %
\DeclareFieldFormat[article]{pages}{#1}             %
\DeclareFieldFormat[article]{year}{(#1)}

\renewbibmacro*{journal+issuetitle}{%
  \setunit{\addcomma\space \addspace}%
  \printfield{journaltitle} %
\newunit}

\DeclareBibliographyDriver{article}{%
  \usebibmacro{bibindex}%
  \usebibmacro{begentry}%
  \usebibmacro{author/translator+others}%
  \setunit{\addcomma\space }\newblock %
  \setunit{\addcomma\space} %
  \printfield{title}
  \newunit{\adddot\space } %
  \printfield{journaltitle}
  \setunit{}\newblock  
  \printfield{year}
  \setunit{\addspace} %
  \iffieldundef{volume}{
  }
  {\printfield{volume}
  \setunit{\addcolon}}
  \printfield{number}
  \setunit{\addcolon}
  \printfield{pages}
  \newunit
  \usebibmacro{finentry}}

\renewbibmacro*{book+year}{%
    \setunit*{\space}%
    \iffieldundef{book}
      {\usebibmacro{date}}
      {\printfield{issue}%
       \setunit*{\addspace}%
       \usebibmacro{date}}%
\newunit}

\DeclareFieldFormat[book]{title}{\mkbibemph{#1}}  %
\DeclareListFormat[book]{publisher}{#1}  %
\DeclareFieldFormat[book]{series}{\mkbibemph{#1}}  %

\DeclareBibliographyDriver{book}{%
  \usebibmacro{bibindex}%
  \usebibmacro{begentry}%
  \newunit\newblock
  \printnames{author}
  \setunit{\adddot\space \addcomma\space } %
  \usebibmacro{byauthor}%
  \newunit
  \setunit{\adddot\space } %
  \usebibmacro{title}%
  \iffieldundef{series}
  {\skipentry}
  {\printfield{series}}
  \newunit
  \setunit{\adddot\space } %
  \printlist{publisher} %
  \usebibmacro*{book+year}
  \usebibmacro{finentry}
}
  
\usepackage{color}
\usepackage{xcolor}
\usepackage{transparent} %

\definecolor{dark_blue}{RGB}{0,9,129}
\definecolor{dark_green}{RGB}{18,172,88} 
\definecolor{dark_blue_cite}{RGB}{0,84,168}
\definecolor{light_grey}{RGB}{190,190,190}

\definecolor{RedDark}{RGB}{139,0,0}
\definecolor{Red}{RGB}{255,0,0}
\definecolor{GreenDark}{RGB}{34,139,34} %
\definecolor{Green}{RGB}{50,205,50} %

\definecolor{Blueone}{RGB}{0,96,173} %
\definecolor{Redone}{RGB}{173,77,0} %

\usepackage{xifthen}%
\usepackage{xparse}
\RequirePackage{physics}
\newcommand{\vari}[1]{\widetilde{#1}}

\newcommand{\fontscal}[1]{#1} %
\newcommand{\fontscalemp}[1]{\mathsf{#1}} %
\newcommand{\fontvec}[1]{\bs{\mathrm{#1}}} %
\newcommand{\fontmat}[1]{\bs{\mathcal{#1}}} %
\DeclareRobustCommand{\time}[1][Default]{
    \IfEqCase{#1}{%
    {Default}{\ensuremath{t}}
        {sim}{\ensuremath{t_{sim}}}%
        {conv}{\ensuremath{t_{conv}}}%
    }[\PackageError{\time}{Undefined option to time: #1}{}]%
}%
\newcommand{\lenLC}{L_{lc}} %

\DeclareRobustCommand{\card}[1][Default]{
    \IfEqCase{#1}{%
      {Default}{\ensuremath{N}}%
        {spe}{\ensuremath{m}}%
        {fluid}{\ensuremath{n_{f}}}%
        {fspe}{\ensuremath{m_k}}%
        {lfspe}{\ensuremath{m_l}}%
        {k}{\ensuremath{k}}%
        {k}{\ensuremath{k^{\prime}}}%
        {j}{\ensuremath{j}}%
        {cell}{\ensuremath{N_{cell}}}%
        {face}{\ensuremath{N_{face}}}%
        {microstate}{\ensuremath{\Omega}}%
        {class}{\ensuremath{n_{c}}}%
    }[\PackageError{card}{Undefined option to cardinal: #1}{}]%
}%
\DeclareRobustCommand{\ite}[1][Default]{
    \IfEqCase{#1}{%
      {Default}{\ensuremath{N}}%
      {spe}{\ensuremath{j}}%
      {speall}{\ensuremath{(j)}}%
      {fspeall}{\ensuremath{(j,k)}}%
      {lfspeall}{\ensuremath{(j,l)}}%
      {lfall}{\ensuremath{(l)}}%
      {fall}{\ensuremath{(k)}}%
      {diff}{\ensuremath{d}}%
      {tspe}{\ensuremath{z}}%
      {fspe}{\ensuremath{j,k}}%
      {fspei}{\ensuremath{i,k}}%
      {lspe}{\ensuremath{j,l}}%
      {fnspe}{\ensuremath{\card[fspe],k}}%
      {f1spe}{\ensuremath{1,k}}%
      {f1spe}{\ensuremath{1,k}}%
      {1spe}{\ensuremath{j,1}}%
      {2spe}{\ensuremath{j,2}}%
      {nspe}{\ensuremath{\card[spe]}}%
      {fluid}{\ensuremath{k}}%
      {fluidp}{\ensuremath{k^{\prime}}}%
      {lfluid}{\ensuremath{l}}%
      {mfluid}{\ensuremath{m}}%
      {fluidl}{\ensuremath{l}}%
      {nfluid}{\ensuremath{\card[fluid]}}%
      {H2O}{\ensuremath{H_{2}O}}%
      {sk}{\ensuremath{k}}%
      {s1}{\ensuremath{1}}%
      {s2}{\ensuremath{2}}%
      {sint}{\ensuremath{\mathrm{I}}}%
      {int}{\ensuremath{\mathrm{I}}}%
      {var}{\ensuremath{\lambda}}%
      {left}{\ensuremath{L}}%
      {right}{\ensuremath{R}}%
      {middle}{\ensuremath{M}}%
      {normal}{\ensuremath{n}}%
      {cap}{\ensuremath{c}}%
      {pm}{\ensuremath{\pm}}%
      {plus}{\ensuremath{-}}%
      {minus}{\ensuremath{+}}%
      {cell}{j}%
      {face}{i}%
    }[\ensuremath{#1}]%
}%

\NewDocumentCommand{\setvar}{o o }{%
  \IfNoValueTF{#1}
    {\ensuremath{\xi}} %
    {%
        \IfNoValueTF{#2}
          {\IfEqCase{#1}{
              {entropy}{\ensuremath{\eta}}%
              {entropyvar}{\ensuremath{\eta_{\lambda}}}%
              {natural}{\ensuremath{\zeta}}%
              {naturalk}{\ensuremath{\zeta_{k}}}%
              {pressure}{\ensuremath{\pi}}%
              {pressurek}{\ensuremath{\pi_{k}}}%
              {thermo}{\ensuremath{\xi}}%
              {thermok}{\ensuremath{\xi_{k}}}%
              {thermoext}{\ensuremath{\Xi_{k}}}%
              {pdf}{\ensuremath{\xi_{f}}}%
          }[\PackageError{\setvar}{Undefined option to setvar: #1}{}]
          }%
        {\IfEqCase{#1}{
            {natural}{\IfEqCase{#2}{%
            {var}{\ensuremath{\zeta^{\tau}}}%
            {vart}{\ensuremath{\zeta^{\tau '}}}%
            {vartt}{\ensuremath{\zeta^{\tau ''}}}%
            	{fluidiso}{\ensuremath{\overline{\zeta_{\ite[fluid]}}}}%
            	}[\ensuremath{\zeta_{\ite[#2]}}]%
            	}
            {pressure}{\IfEqCase{#2}{%
            	{fluidiso}{\ensuremath{\overline{\pi_{\ite[fluid]}}}}%
            	}[\ensuremath{\pi_{\ite[#2]}}]%
				}%
            {thermo}{\IfEqCase{#2}{%
            	{fluidiso}{\ensuremath{\overline{\xi_{\ite[fluid]}}}}%
            	}[\ensuremath{\xi_{\ite[#2]}}]%
            	}%
            {thermoext}{\ensuremath{\Xi_{\ite[#2]}}}%
          }
        }
      }
}

\NewDocumentCommand{\speed}{o o o }{%
  \IfNoValueTF{#1}
    {\ensuremath{\bs{v}}}%
    {%
        \IfNoValueTF{#2}%
          {\IfEqCase{#1}{%
        {Default}{\ensuremath{\bs{v}}}%
        {vk}{\bs{v}_{k}}%
        {sk}{v_{k}}%
        {sint}{v_{I}}%
        {surface}{\fontvec{v}_{s}}%
        {int}{\bs{v}_{I}}%
        {moy}{\overline{\bs{v}}}%
        {delta}{\delta \bs{v}}%
        {diff}{\bs{v}_{d}}%
        {mean}{\bs{v}_{h}}%
        {v}{\bs{v}}%
        {v1}{\bs{v}_{1}}%
        {v2}{\bs{v}_{2}}%
        {x}{v_{x}}%
        {y}{v_{y}}%
        {z}{v_{z}}%
        {s}{v}%
        {sx1}{v_{x_{1}}}%
        {sx2}{v_{x_{2}}}%
        {sx3}{v_{x_{3}}}%
        {sxi}{v_{x_{i}}}%
        {sdiff}{v_{d}}%
        {s1}{v_{1}}%
        {s2}{v_{2}}%
        {sair}{V_{air}}%
        {sH2O}{V_{H_{2}O}}%
        {sl}{V_{L}}%
        {sg}{V_{G}}%
        {sound}{a}%
        {sound1}{a_{1}}%
        {sound2}{a_{2}}%
        {var}{\ensuremath{\bs{v}_{\lambda}}}%
        {normal}{\ensuremath{v_{\ite[normal]}}}%
          }[\ensuremath{\bs{v}_{\ite[#1]}}]%
          }%
          {\IfNoValueTF{#3}
            {\IfEqCase{#1}{%
            {wave}{\ensuremath{S_{\ite[#2]}}}%
            {normal}{\ensuremath{v_{n,\ite[#2]}}}%
            {sound}{\ensuremath{a_{\ite[#2]}}}%
            {v}{\ensuremath{\bs{v}_{\ite[#2]}}}%
            {s}{\IfEqCase{#2}{%
            		{fluctu}{\ensuremath{v^{\prime}}}%
            		}[\ensuremath{v_{\ite[#2]}}]}%
            {S}{\IfEqCase{#2}{%
            		{fluctu}{\ensuremath{V^{\prime}}}%
            		}[\ensuremath{V_{\ite[#2]}}]}%
            {i}{\IfEqCase{#2}{%
            		{fluctu}{\ensuremath{v^{\prime}_{i}}}%
            		}[\PackageError{speed}{Undefined option to speed: 0  = #2}{}]}%
            {j}{\IfEqCase{#2}{%
            		{fluctu}{\ensuremath{v^{\prime}_{j}}}%
            		}[\PackageError{speed}{Undefined option to speed: 0  = #2}{}]}%
            {x}{\IfEqCase{#2}{%
            		{moy}{\ensuremath{V}}%
            		{delta}{\ensuremath{ v_{x}}}%
            		{fluctu}{\ensuremath{v^{\prime}_{x}}}%
            		}[\PackageError{speed}{Undefined option to speed: 0  = #2}{}]}%
            {y}{\IfEqCase{#2}{%
            		{delta}{\ensuremath{ v_{y}}}%
            		{fluctu}{\ensuremath{v^{\prime}_{y}}}%
            		}[\PackageError{speed}{Undefined option to speed: 0  = #2}{}]}%
            {z}{\IfEqCase{#2}{%
            		{delta}{\ensuremath{ v_{z}}}%
            		{fluctu}{\ensuremath{v^{\prime}_{z}}}%
            		}[\PackageError{speed}{Undefined option to speed: 0  = #2}{}]}%
            {fluid}{\ensuremath{\qty(\bs{v}_{\ite[#1]})_{\ite[#2]}}}%
            {int}{\ensuremath{\bs{v}_{\ite[#1],\ite[#2]}}}%
            }[\PackageError{speed}{Undefined option to speed: #2}{}]%
            }%
            {\IfEqCase{#1}{%
            	  {x}{\IfEqCase{#2}{%
            		{moy}{\IfEqCase{#3}{%
            			{cst}{\ensuremath{V}_{0}}%
            			}[\PackageError{speed}{Undefined option to speed: 0  = #3}{}]}%
            		}[\PackageError{speed}{Undefined option to speed: 0  = #2}{}]}%
              {vol}{\ensuremath{V_{\ite[#2],\ite[#3]}}}%
              {spec}{\ensuremath{v_{\ite[#2],\ite[#3]}}}%
              {frac}{\ensuremath{\alpha_{\ite[#2],\ite[#3]}}}%
			 {sound}{\IfEqCase{#2}{%
                          {fluid}{\IfEqCase{#3}{%
                            {K}{(a_{\ite[#2]})_{\ite[#3]}}%
                          }%
                        }%
                      }%
                  }%
              }[\PackageError{speed}{Undefined option with 3 entries to speed: #1}{}]
            }%
          }%
    }%
}
\NewDocumentCommand{\pressure}{o o }{%
  \IfNoValueTF{#1}
    {\ensuremath{\ensuremath{p}}} %
    {%
        \IfNoValueTF{#2}
          {\IfEqCase{#1}{
        	{fluid}{p_{\ite[fluid]}}%
        	{var}{\ensuremath{p_{\lambda}}}%
        	{delta}{\ensuremath{\delta p}}%
        	{moy}{\ensuremath{\overline{p}}}%
        	{smel}{\ensuremath{\Delta p}}%
        	{extend}{\ensuremath{\mathcal{P}}}%
        	{contact}{\ensuremath{p_{c}}}%
          }[\ensuremath{p_{\ite[#1]}}]
          }%
          {\IfEqCase{#1}{%
        		{fluid}{\IfEqCase{#2}{%
        			{PG}{\ensuremath{p^{PG}_{\ite[fluid]}}}%
        			{iso}{\ensuremath{p^{o}_{\ite[fluid]}}}%
        			}[\ensuremath{\qty(p_{\ite[fluid]})_{\ite[#2]}}]%
        			}
        		{int}{\ensuremath{p_{\ite[#1],\ite[#2]}}}%
          	{var}{\ensuremath{p_{\ite[#2], \ite[var]}}}%
       	 	{delta}{\ensuremath{\delta p_{\ite[#2]}}}%
          	{extend}{\IfEqCase{#2}{%
          		{delta}{\delta \mathcal{P}}%
          		}[\ensuremath{\mathcal{P}_{\ite[#2]}}]%
          	}%
          	{smel}{\ensuremath{\Delta p_{\ite[#2]}}}%
          	}[\PackageError{pressure}{Undefined option with 2 entries to pressure: #1}{}]
          }
    }
}
\NewDocumentCommand{\density}{o o }{%
  \IfNoValueTF{#1}
    {\ensuremath{\ensuremath{\rho}}} %
    {%
        \IfNoValueTF{#2}
          {\IfEqCase{#1}{
        	{fluid}{\rho_{\ite[fluid]}}%
        	{var}{\ensuremath{\rho_{\lambda}}}%
        	{delta}{\ensuremath{\delta \rho}}%
        	{moy}{\ensuremath{\overline{\rho}}}%
          }[\ensuremath{\rho_{\ite[#1]}}]
          }%
          {\IfEqCase{#1}{
          {fluid}
            {\IfEqCase{#2}{
            	{moy}{\ensuremath{\overline{\rho}_{\ite[#1]}}}%
            	}[\ensuremath{\qty(\rho_{\ite[fluid]})_{\ite[#2]}}]
            	}%
            {var}{\ensuremath{\rho_{\ite[#2], \ite[var]}}}%
            }[\PackageError{density}{Undefined option with 2 entries to density: #1}{}]
          }
    }
}
\DeclareRobustCommand{\volume}[1][Default]{
    \IfEqCase{#1}{%
    {Default}{\ensuremath{V}}%
    {sk}{\ensuremath{V_{k}}}%
    {s1}{\ensuremath{V_{1}}}%
    {s2}{\ensuremath{V_{2}}}%
    {sp}{\ensuremath{V^{\prime}}}%
    {fluid}{\ensuremath{V_{\ite[fluid]}}}%
    {cell}{\ensuremath{\mathcal{V}_{\ite[cell]}}}%
    }[\PackageError{volume}{Undefined option to volume: #1}{}]%
}%

\DeclareRobustCommand{\volspec}[1][Default]{
    \IfEqCase{#1}{%
    {Default}{\ensuremath{v}}%
    {sk}{\ensuremath{v_{k}}}%
    {fluid}{\ensuremath{v_{\ite[fluid]}}}%
    {s1}{\ensuremath{v_{1}}}%
    {s2}{\ensuremath{v_{2}}}%
    {sp}{\ensuremath{v^{\prime}}}%
    }[\PackageError{volspec}{Undefined option to vol spec: #1}{}]%
}%

\NewDocumentCommand{\vol}{o o o }{%
  \IfNoValueTF{#1}
    {\ensuremath{V}} %
    {%
        \IfNoValueTF{#2}
          {\IfEqCase{#1}{
            {vol}{\ensuremath{V}}   %
            {fluid}{\ensuremath{V_{\ite[fluid]}}}   %
            {fluidall}{\ensuremath{V_{(\ite[fluid])}}}    %
              {spec}{\ensuremath{v}}    %
              {specp}{\ensuremath{v^{\prime}}}    %
              {specpp}{\ensuremath{v^{\prime \prime}}}    %
              {frac}{\ensuremath{\alpha}} %
              {mel}{\ensuremath{\Delta V}}%
          }[\PackageError{vol}{Undefined option to volume: #1}{}]
          }%
          {\IfNoValueTF{#3}
            {\IfEqCase{#1}{
              {vol}{\ensuremath{V_{\ite[#2]}}}
                {frac}{\ensuremath{\alpha_{\ite[#2]}}}%
                {fracall}{\ensuremath{\alpha_{(\ite[#2])}}}%
                {fract}{\ensuremath{\tilde{\alpha}_{\ite[#2]}}}%
                {spec}{\ensuremath{v_{\ite[#2]}}}%
            }[\PackageError{vol}{Undefined option to volume: #2}{}]
            }%
            {\IfEqCase{#1}{
              {vol}{\ensuremath{V_{\ite[#2],\ite[#3]}}}   %
                {spec}{\ensuremath{v_{\ite[#2],\ite[#3]}}}
                {frac}{\IfEqCase{#3}{%
                		{fluctu}{\ensuremath{\alpha_{\ite[#2]}^{\prime}}}%
                		}[\ensuremath{\alpha_{\ite[#2],\ite[#3]}}]}%
                }[\PackageError{vol}{Undefined option with 3 entries to volume: #1}{}]
            }
          }
    }
}

\NewDocumentCommand{\mass}{o o o }{%
  \IfNoValueTF{#1}
    {\ensuremath{M}} %
    {%
        \IfNoValueTF{#2}
          {\IfEqCase{#1}{
            {vol}{\ensuremath{M}}
            {fluid}{\ensuremath{M_{\ite[fluid]}}}
            {fluidall}{\ensuremath{M_{(\ite[fluid])}}}
            {fspeall}{\ensuremath{M_{(\ite[spe],\ite[fluid])}}}
            {frac}{\ensuremath{Y}}%
            {sk}{\ensuremath{M_{k}}}%
            {transfer}{\ensuremath{\Gamma}}%
          }[\PackageError{mass}{Undefined option to mass: #1}{}]
          }%
          {\IfNoValueTF{#3}
            {\IfEqCase{#1}{
              {vol}{\ensuremath{M_{\ite[#2]}}}
              {frac}{\IfEqCase{#2}{%
              	{moy}{\ensuremath{\overline{Y}}}%
              	{delta}{\ensuremath{\delta Y}}%
              	}[\ensuremath{\ensuremath{y_{\ite[#2]}}}]}%
              {fracb}{\ensuremath{\overline{y}_{\ite[#2]}}}%
              {fract}{\ensuremath{\tilde{y}_{\ite[#2]}}}%
              {fracall}{\ensuremath{y_{(\ite[#2])}}}%
            {transfer}{\ensuremath{\Gamma_{\ite[#2]}}}%
            }[\PackageError{mass}{Undefined option to mass: #2}{}]
            }%
            {\IfEqCase{#1}{
              {vol}{\ensuremath{M_{\ite[#2],\ite[#3]}}}   %
                {frac}{\ensuremath{y_{\ite[#2],\ite[#3]}}}
                {fracb}{\ensuremath{\overline{y}_{\ite[#2],\ite[#3]}}}%
                }[\PackageError{mass}{Undefined option with 3 entries to mass: #1}{}]
            }
          }
    }
}

\DeclareRobustCommand{\massfrac}[1][Default]{
    \IfEqCase{#1}{%
        {Default}{\ensuremath{Y}}%
        {sk}{\ensuremath{y_{k}}}%
        {s}{\ensuremath{y}}%
        {s1}{\ensuremath{y_{1}}}%
        {s2}{\ensuremath{y_{2}}}%
        {sN}{\ensuremath{y_{N}}}%
        {sNj}{\ensuremath{y_{N_{j}}}}%
        {var}{\ensuremath{\massfrac_{\lambda}}}
        {skp}{\ensuremath{y_{k^{\prime}}}}%
    }[\PackageError{massfrac}{Undefined option to massfrac: #1}{}]%
}%

\NewDocumentCommand{\tem}{o o }{%
  \IfNoValueTF{#1}
    {\ensuremath{T}} %
    {\IfNoValueTF{#2}
  		{\IfEqCase{#1}{
		    	{fluid}{\ensuremath{T_{\ite[fluid]}}}%
		    	{int}{\ensuremath{T_{\ite[int]}}}%
	    }[\ensuremath{T_{\ite[#1]}}]%
        }%
        {\IfEqCase{#1}{
          {fluid}{\IfEqCase{#2}{
          	{fluidp}{\ensuremath{T_{\ite[#1],\ite[#2]}}}%
          	{lfluid}{\ensuremath{T_{\ite[#1],\ite[#2]}}}%
          		}[\ensuremath{T_{\ite[#1],\ite[#2]}}]%
          	}%
          }
        }
      }
}

\NewDocumentCommand{\gibbs}{o o }{%
  \IfNoValueTF{#1}
    {\ensuremath{g}} %
    {\IfNoValueTF{#2}
  		{\IfEqCase{#1}{
      		{smel}{\ensuremath{\Delta g}}%
      		{mel}{\ensuremath{\Delta g}}%
      	}[\ensuremath{g_{\ite[#1]}}]%
        }%
        {\IfEqCase{#1}{
          {fspe}{\IfEqCase{#2}{%
          	{iso}{\ensuremath{g^{o}_{\ite[fspe]}}}%
          		}[\PackageError{gibbs}{Undefined option to gibbs fspe iso: #2}{}]%
          	}%
          {fluid}{\IfEqCase{#2}{
          	{PG}{\ensuremath{g^{PG}_{\ite[fluid]}}}%
          	{iso}{\ensuremath{g^{o}_{\ite[fluid]}}}%
          		}[\ensuremath{g_{\ite[#2]}}]%
          	}%
          {mel}{\ensuremath{\Delta g_{\ite[#2]}}}%
          {smel}{\ensuremath{\Delta g_{\ite[#2]}}}%
          {mix}{\ensuremath{G_{\ite[#2]}}}%
          }
        }
      }
}

\NewDocumentCommand{\entropy}{o o }{%
  \IfNoValueTF{#1}
    {\ensuremath{s}} %
    {%
        \IfNoValueTF{#2}
          {\IfEqCase{#1}{
            {flux}{\ensuremath{\fontvec{G}}}%
            {vol}{\ensuremath{S}}%
              {fluid}{\ensuremath{s_{\ite[fluid]}}}%
              {sk}{\ensuremath{s_{k}}}%
              {svolk}{\ensuremath{S_{k}}}%
              {svol}{\ensuremath{S}}%
              {s}{\ensuremath{s}}%
              {s1}{\ensuremath{s_{1}}}%
              {s2}{\ensuremath{s_{2}}}%
              {smel}{\ensuremath{\Delta s}}%
              {PG}{\ensuremath{s^{PG}}}%
          }[\PackageError{entropy}{Undefined option to entropy: #1}{}]
          }%
        {\IfEqCase{#1}{
          {fluid}{\IfEqCase{#2}{
          	{PG}{\ensuremath{s_{\ite[fluid]}^{PG}}}%
          	{iso}{\ensuremath{s_{\ite[fluid]}^{o}}}%
          	}[\ensuremath{\qty(s_{\ite[fluid]})_{\ite[#2]}}]%
          	}%
          {vol}{\ensuremath{S_{\ite[#2]}}}%
          {frac}{\ensuremath{s_{\ite[#2]}}}%
          {var}{\ensuremath{s_{\ite[#2], \ite[var]}}}%
          {smel}{\ensuremath{\Delta s_{\ite[#2]}}}%
          }
        }
      }
}
\NewDocumentCommand{\enthalpy}{o o }{%
  \IfNoValueTF{#1}
    {\ensuremath{s}} %
    {%
        \IfNoValueTF{#2}
          {\IfEqCase{#1}{
            {vol}{\ensuremath{H}}%
              {fluid}{\ensuremath{h_{\ite[fluid]}}}%
              {sk}{\ensuremath{h_{k}}}%
              {svolk}{\ensuremath{H_{k}}}%
              {svol}{\ensuremath{H}}%
              {s}{\ensuremath{h}}%
              {s1}{\ensuremath{h_{1}}}%
              {s2}{\ensuremath{h_{2}}}%
              {smel}{\ensuremath{\Delta h}}%
          }[\ensuremath{h_{\ite[#1]}}]
          }%
        {\IfEqCase{#1}{
          {fluid}{\ensuremath{\qty(h_{\ite[fluid]})_{\ite[#2]}}}%
          {vol}{\ensuremath{H_{\ite[#2]}}}%
          {frac}{\ensuremath{h_{\ite[#2]}}}%
          {var}{\ensuremath{h_{\ite[#2], \ite[var]}}}%
          }
        }
      }
}

\NewDocumentCommand{\fpot}{o o o }{%
  \IfNoValueTF{#1}
    {\ensuremath{f}} %
    {%
        \IfNoValueTF{#2}
          {\IfEqCase{#1}{%
            {volfrac}{\ensuremath{f_{\alpha}}}%
            {interf}{\ensuremath{F^{\Sigma}}}%
          	}[\PackageError{fpot}{Undefined option to function potential: #1}{}]
          }%
          {\IfNoValueTF{#3}
            {\IfEqCase{#1}{%
         	 	{volfrac}{\ensuremath{f_{\alpha_{\ite[#2]}}}}%
            		{interf}{\ensuremath{f^{\Sigma}_{\ite[#2]}}}%
          		}[\PackageError{fpot}{Undefined option to fpot: #2}{}]%
            }%
            {\IfEqCase{#1}{%
          		{volfrac}{\IfEqCase{#2}{%
            			{mix}{\ensuremath{F_{\alpha_{\ite[#3]}}}}%
            			{mel}{\ensuremath{\Delta f_{\alpha_{\ite[#3]}}}}%
          			}[\PackageError{fpot}{Undefined option to fpot: #1, #2, #3 with volfrac}{}]
          			}
            		{interf}{\ensuremath{F^{\Sigma}_{\ite[#2],\ite[#3]}}}%
              }[\PackageError{fpot}{Undefined option with 3 entries to fpot: #1}{}]
            }
          }
    }
}
\NewDocumentCommand{\energy}{o o }{%
  \IfNoValueTF{#1}
    {\ensuremath{e}} %
    {%
        \IfNoValueTF{#2}
          {\IfEqCase{#1}{
            {vol}{\ensuremath{E}}
            {svol}{\ensuremath{E}}
            {frac}{\ensuremath{e}}%
            {sk}{\ensuremath{e_{k}}}%
            {fluid}{\ensuremath{e_{\ite[fluid]}}}%
            {fluidp}{\ensuremath{e_{\ite[fluidp]}}}%
            {s1}{\ensuremath{e_{1}}}%
            {s2}{\ensuremath{e_{2}}}%
              {smel}{\ensuremath{\Delta e}}%
              {PG}{\ensuremath{e^{PG}}}%
              {tot}{\ensuremath{E}}%
              {diff}{\ensuremath{E_{d}}}%
              {kinetic}{\ensuremath{\mathscr{K}}}%
              {internal}{\ensuremath{\epsilon}}%
              {kineticsmall}{\ensuremath{\mathscr{K}_i}}%
              {interfacial}{\ensuremath{\mathscr{U}_i}}%
              {potential}{\ensuremath{\mathscr{U}}}%
              {capl}{\ensuremath{\mathscr{E}_{c}^{l}}}%
              {caps}{\ensuremath{\mathscr{E}_{c}^{s}}}%
              {free}{\ensuremath{f}}%
          }[\ensuremath{e_{\ite[#1]}}]%
          }%
        {\IfEqCase{#1}{
          {fluid}{\IfEqCase{#2}{%
          	{PG}{\ensuremath{e_{\ite[fluid]}^{PG}}}%
          	{iso}{\ensuremath{e_{\ite[fluid]}^{o}}}%
          	}[\ensuremath{\qty(e_{\ite[fluid]}_{\ite[#2]}}]%
          	}%
          {vol}{\ensuremath{E_{\ite[#2]}}}
          {frac}{\ensuremath{e_{\ite[#2]}}}%
          {tot}{\ensuremath{E_{\ite[#2]}}}%
          {internal}{\ensuremath{\epsilon_{\ite[#2]}}}%
          {free}{\ensuremath{f}_{\ite[#2]}}%
          {smel}{\ensuremath{\Delta e_{\ite[#2]}}}%
          }
        }
      }
}
\NewDocumentCommand{\force}{o o }{%
  \IfNoValueTF{#1}
    {\ensuremath{\bs{f}}} %
    {%
        \IfNoValueTF{#2}
          {\IfEqCase{#1}{
            {gravity}{\ensuremath{\bs{g}}}%
            {body}{\ensuremath{\bs{f}^{b}}}%
            {surface}{\ensuremath{\bs{f}^{s}}}%
            {drag}{\ensuremath{\bs{F}_{d}}}%
          }[\PackageError{force}{Undefined option to force: #1}{}]
          }%
        {\IfEqCase{#1}{
          {gravity}{\ensuremath{\bs{g}_{\ite[#2]}}}%
          {body}{\ensuremath{\bs{f}^{b}_{\ite[#2]}}}%
          {drag}{\ensuremath{\bs{F}_{d,\ite[#2]}}}%
          }
        }
      }
}
\NewDocumentCommand{\heat}{o o }{%
  \IfNoValueTF{#1}
    {\ensuremath{\bs{q}}}%
    {%
        \IfNoValueTF{#2}
          {\IfEqCase{#1}{
            {flux}{\ensuremath{\bs{q}}}%
            {source}{\ensuremath{\bs{\dot{q}}}}%
          }[\PackageError{heat}{Undefined option to heat: #1}{}]
          }%
        {\IfEqCase{#1}{
          {flux}{\ensuremath{\bs{q}_{\ite[#2]}}}
          {source}{\ensuremath{\bs{\dot{q}}_{\ite[#2]}}}
         }
        }
      }
}
\newcommand{\Srelax}{\fontvec{r}}     %
\newcommand{\Sstrain}{\fontvec{T}}    %

\NewDocumentCommand{\coeff}{o o }{%
  \IfNoValueTF{#1}
    {\ensuremath{\lambda}}%
    {%
        \IfNoValueTF{#2} %
          {\IfEqCase{#1}{
            {relax}{\ensuremath{\lambda}}%
            {surften}{\ensuremath{\sigma}}%
          }[\PackageError{coeff}{Undefined option to coeff: #1}{}]
          }%
        {%
        \IfEqCase{#1}{%
          {relax}{\IfEqCase{#2}{%
      			{puls}{\ensuremath{\epsilon}}%
      			{speed}{\ensuremath{\lambda}}%
      			{pressure}{\ensuremath{\mu}}%
          		}[\PackageError{coeff}{Undefined option to coeff - relax: #2}{}]}         		
        }[\PackageError{coeff}{Undefined option to coeff #1 with a second #2}{}]
        }%
     }%
}%

\NewDocumentCommand{\source}{o o o }{%
  \IfNoValueTF{#1}
    {\ensuremath{\mathcal{T}}}%
    {%
        \IfNoValueTF{#2}
          {\IfEqCase{#1}{
            {stress}{\ensuremath{\fontvec{T}}}%
            {viscous}{\ensuremath{\fontvec{D}}}%
            {relax}{\ensuremath{\fontvec{r}}}%
            {interfarea}{\ensuremath{\fontvec{S}_{\Sigma}}}%
            {meancurv}{\ensuremath{\fontvec{S}_{\meancurv}}}%
            {gausscurv}{\ensuremath{\fontvec{S}_{\gausscurv}}}%
           }[\PackageError{source}{Undefined option to source: #1}{}]%
          }%
          {\IfNoValueTF{#3}
            {\IfEqCase{#1}{%
          		{stress}{\ensuremath{\fontvec{T}_{\ite[#2]}}}%
          		{viscous}{\ensuremath{\fontvec{D}_{\ite[#2]}}}%
          		{relax}{\IfEqCase{#2}{%
            			{speed}{\ensuremath{\fontvec{r}^{\speed[s]}}}%
            			{pressure}{\ensuremath{\fontvec{r}^{\pressure}}}%
          			}[\PackageError{source}{Undefined option to source: #1, #2 with relax}{}]%
          			}%
          		}[\PackageError{source}{Undefined option to source: #2}{}]%
            }%
            {\IfEqCase{#1}{%
          		{relax}{\IfEqCase{#2}{
            			{speed}{\ensuremath{\fontvec{r}^{\speed[s]}_{\ite[#3]}}}%
            			{pressure}{\ensuremath{\fontvec{r}^{\pressure}_{\ite[#3]}}}%
          			}[\PackageError{source}{Undefined option to source: #1, #2, #3 with relax}{}]
          			}
              }[\PackageError{speed}{Undefined option with 3 entries to speed: #1}{}]
            }
          }
    }
}
\NewDocumentCommand{\exchange}{o o }{%
  \IfNoValueTF{#1}
    {\ensuremath{m}}%
    {%
        \IfNoValueTF{#2}
          {\IfEqCase{#1}{
            {mass}{\ensuremath{c^{+}}}%
            {momentum}{\ensuremath{\fontvec{m^{+}}}}%
            {energy}{\ensuremath{e^{+}}}%
          }[\PackageError{source}{Undefined option to source: #1}{}]
          }%
        {\IfEqCase{#1}{
            {mass}{\ensuremath{c^{+}_{\ite[#2]}}}%
            {momentum}{\ensuremath{\fontvec{m^{+}}_{\ite[#2]}}}%
            {energy}{\ensuremath{e^{+}_{\ite[#2]}}}%
         }
        }
      }
}

\DeclareRobustCommand{\rhovar}[1][Default]{
    \IfEqCase{#1}{%
        {Default}{\vari{\rho}}%
    }[\PackageError{rhovar}{Undefined option to rhovar: #1}{}]%
}%

\NewDocumentCommand{\pulsation}{o }{%
  \IfNoValueTF{#1}
    {\ensuremath{\omega}} %
  {\IfEqCase{#1}{
      {suppl}{\ensuremath{\Delta \omega}}%
      {delta}{\ensuremath{\delta \omega}}%
	  {interfarea}{\ensuremath{\sigma}}%
      }
      [\ensuremath{\omega_{\ite[#1]}}]%
    }
}

\NewDocumentCommand{\Lagr}{o o }{%
  \IfNoValueTF{#1}
    {\ensuremath{L}} %
    {%
        \IfNoValueTF{#2}
          {\IfEqCase{#1}{
      		{filter}{\ensuremath{L_{c}}}%
      		{star}{\ensuremath{L^{\star}}}%
      		{var}{\accentset{\circ}{\ensuremath{L}}}%
      		{fluid}{\ensuremath{L_{\ite[fluid]}}}%
          }[\ensuremath{L_{\ite[#1]}}]%
          }%
        {\IfEqCase{#1}{
             {star}{\ensuremath{L^{\star}_{\ite[#2]}}}%
        		}[\PackageError{Lagr}{Undefined option to Lagrangian #1 + #2}{}]%
      	}
     }
}
\DeclareRobustCommand{\Lagrdiff}[1][Default]{
    \IfEqCase{#1}{%
      {Default}{\ensuremath{\dd L}}%
      {u}{\ensuremath{\bs{K}}}%
      {gradalpha}{\ensuremath{\bs{D}}}%
      {dtalpha}{\ensuremath{M}}%
      {dtsigma}{\ensuremath{N}}%
    }[\PackageError{Lagrdiff}{Undefined option to Lagrdiff: #1}{}]%
}%

\newcommand{\leftdiv}[0]{\ensuremath{\bs{\, \cdot \, \nabla}}}

\NewDocumentCommand{\LAPvar}{o }{%
  \IfNoValueTF{#1}%
    {\ensuremath{\lambda}}%
  {\IfEqCase{#1}{%
        {0}{\ensuremath{\lambda}}%
        {1}{\ensuremath{\lambda_{1}}}%
        {2}{\ensuremath{\lambda_{2}}}%
        {kp}{\lambda_{k^{\prime}}}%
        {fluidp}{\lambda_{\ite[fluidp]}}%
		{t}{\ensuremath{\lambda}}%
		{vect}{\bs{\chi}} %
		{fluid}{\lambda_{\ite[fluid]}} %
      }[\ensuremath{\lambda_{\ite[#1]}}]%
    }%
}%

\NewDocumentCommand{\pathM}{o o }{%
  \IfNoValueTF{#1}
    {\ensuremath{\gamma}} %
    {%
        \IfNoValueTF{#2}
          {\IfEqCase{#1}{
             {var}{\ensuremath{\Gamma}}%
        		{0}{\ensuremath{\gamma_{\lambda}}}%
        		{1}{\ensuremath{\gamma_{\lambda_{1}}}}%
        		{2}{\ensuremath{\gamma_{\lambda_{2}}}}%
        		{kp}{\gamma_{\lambda_{k^{\prime}}}}%
        		{fluidp}{\gamma_{\lambda_{\ite[fluidp]}}}%
			{t}{\gamma_{\ensuremath{\lambda}}}%
			{vect}{\gamma_{\bs{\chi}}} %
			{fluid}{\gamma_{\lambda_{\ite[fluid]}}} %
          }[\ensuremath{\gamma_{\ite[#1]}}]%
          }%
        {\IfEqCase{#1}{
             {var}{\ensuremath{\Gamma_{\ite[#2]}}}%
        		}[\PackageError{pathM}{Undefined option pathM to #1 + #2}{}]%
      	}
     }
}

\NewDocumentCommand{\xlag}{o }{%
  \IfNoValueTF{#1}
    {\ensuremath{\bs{X}}} %
  {\IfEqCase{#1}{%
      {tobedefined}{\ensuremath{\bs{X}}}%
      }%
      [\ensuremath{\bs{X}_{\ite[#1]}}]%
    }%
}%
\NewDocumentCommand{\xeul}{o }{%
  \IfNoValueTF{#1}
    {\ensuremath{\bs{x}}} %
  {\IfEqCase{#1}{%
      {tobedefined}{\ensuremath{\bs{x}}}%
      }%
      [\ensuremath{\bs{x}_{\ite[#1]}}]%
    }%
}%

\NewDocumentCommand{\pathx}{o o o }{%
  \IfNoValueTF{#1}
    {\ensuremath{\varphi}} %
    {%
        \IfNoValueTF{#2}
          {\IfEqCase{#1}{
             {eul}{\ensuremath{\vb*{\varphi}}}%
             {lag}{\ensuremath{\vb*{\varphi}^L}}%
             {var}{\ensuremath{\hat{\varphi}}}%
          }[\ensuremath{\varphi_{\ite[#1]}}]%
          }%
        	  {\IfNoValueTF{#3}
            {\IfEqCase{#1}{
             {var}{\ensuremath{\hat{\varphi}_{\ite[#2]}}}%
             {lag}{\IfEqCase{#2}{
             		{var}{\ensuremath{\vb*{\varphi}^L_{\LAPvar}}}%
             		}[\ensuremath{\vb*{\varphi}^L_{#2}}]%
          		  }%
             {eul}{\IfEqCase{#2}{
             		{var}{\ensuremath{\vb*{\varphi}_{\LAPvar}}}%
             		}[\PackageError{pathx}{Undefined option to #1=eul + #2}{}]%
          		  }%
            }[\PackageError{pathx}{Undefined option to #1 + #2}{}]
            }%
            {\IfEqCase{#1}{
             	{lag}{\IfEqCase{#2}{
             		{var}{\ensuremath{\vb*{\varphi}^L_{\LAPvar[#3]}}}%
             		}[\PackageError{pathx}{Undefined option to #1=lag + #2 + #3}{}]%
          		  }%
                }[\PackageError{pathx}{Undefined option with 3 entries to pathx: #1}{}]
            }
          }
    }
}

\NewDocumentCommand{\SV}{o }{%
  \IfNoValueTF{#1}
    {\ensuremath{\fontvec{u}}} %
  {\IfEqCase{#1}{
      {face}{\ensuremath{\fontvec{u}_{\ite[face]}}}%
      {int}{\ensuremath{\fontvec{u}_{\ite[int]}}}%
      {t}{\ensuremath{\tilde{\fontvec{u}}}}%
      }
      [\ensuremath{\fontvec{u}_{\ite[#1]}}]%
    }
}

\NewDocumentCommand{\CV}{o o }{%
  \IfNoValueTF{#1}
    {\ensuremath{\fontvec{q}}} %
    {%
        \IfNoValueTF{#2}
          {\IfEqCase{#1}{
             {scal}{\ensuremath{q}}%
          }[\ensuremath{\fontvec{q}_{\ite[#1]}}]%
          }%
        {\IfEqCase{#1}{
            {scal}{\ensuremath{q_{\ite[#2]}}}%
          }
        }
      }
}

\NewDocumentCommand{\CF}{o }{%
  \IfNoValueTF{#1}
    {\ensuremath{\fontvec{f}}} %
  {\IfEqCase{#1}{
      {face}{\ensuremath{\fontvec{f}_{\ite[face]}}}%
      {int}{\ensuremath{\fontvec{f}_{\ite[int]}}}%
      }
      [\ensuremath{\fontvec{f}_{\ite[#1]}}]%
    }
}
\NewDocumentCommand{\NCF}{o o }{%
  \IfNoValueTF{#1}
    {\ensuremath{\fontmat{N}}} %
    {%
        \IfNoValueTF{#2}
          {\IfEqCase{#1}{
      		{v}{\ensuremath{\fontvec{n}}}%
          }[\ensuremath{\fontmat{N}_{\ite[#1]}}]%
          }%
        {\IfEqCase{#1}{
            {v}{\ensuremath{\fontvec{n}_{\ite[#2]}}}%
          }
        }
      }
}

\newcommand{\NCFone}{\fontvec{n}_{1}}     
\newcommand{\NCFtwo}{\fontvec{n}_{2}}

\newcommand{\MES}{\fontscalemp{H}}          %
\DeclareRobustCommand{\transfer}[1]{%
    \IfEqCase{#1}{%
        {salpha}{\fontscal{t_{\alpha}}} %
        {s1}{\fontscal{t_{1}}} %
        {s2}{\fontscal{t_{2}}} %
        {sk}{\fontscal{t_{k}}} %
        {v}{\fontvec{t}} %
        {m}{\fontmat{T}}%
    }[\PackageError{transfer}{Undefined option to transfer: #1}{}]%
}%
\newcommand{\x}[1][Default]{
    \IfEqCase{#1}{%
        {Default}{\ensuremath{\bs{x}}} %
    }[\PackageError{x}{Undefined option to x: #1}{}]%
}%

\NewDocumentCommand{\radius}{o }{%
  \IfNoValueTF{#1}
    {\ensuremath{\ensuremath{r}}}%
  {\IfEqCase{#1}{
      {add}{\ensuremath{r_{add}}}%
      }
      [\ensuremath{r_{\ite[#1]}}]%
    }
}
\NewDocumentCommand{\diam}{o }{%
  \IfNoValueTF{#1}
    {\ensuremath{\ensuremath{d}}}%
  {\IfEqCase{#1}{
      {zl}{\ensuremath{Z_{l}}}%
      {zg}{\ensuremath{Z_{g}}}%
      }
      [\ensuremath{d_{\ite[#1]}}]%
    }
}

\newcommand{\charlen}[1][Default]{
    \IfEqCase{#1}{%
        {Default}{\ensuremath{\ell}} %
        {c}{\ensuremath{\ell}_{c}} %
        {filter}{\ensuremath{\ell}_{c}} %
        {DNS}{\ensuremath{\ell_{DNS}}} %
        {mfp}{\ensuremath{\ell_{fr}}} %
        {fp}{\ensuremath{\ell_{p}}} %
        {macro}{\ensuremath{L}} %
        {KH}{\ensuremath{d}} %
    }[\PackageError{charlen}{Undefined option to charlen: #1}{}]%
}%

\NewDocumentCommand{\wave}{o o }{%
  \IfNoValueTF{#1}
    {\ensuremath{wave}} %
    {%
        \IfNoValueTF{#2}
          {\IfEqCase{#1}{
        {len}{\ensuremath{\lambda}}%
        {num}{\ensuremath{k}}%
        {puls}{\ensuremath{n}}%
          }[\ensuremath{\vb{n}_{\ite[#1]}}]
          }%
          {\IfEqCase{#1}{
        {len}{\ensuremath{\lambda}}%
        {num}{\ensuremath{k}}%
        {puls}{\IfEqCase{#2}{%
        		{tot}{\ensuremath{N}}%
          	}[\PackageError{wave}{Undefined option with 2 entries to wave: #1}{}]
          }
        }[\PackageError{wave}{Undefined option with 2 entries to wave: #1}{}]}
    }
}

\newcommand{\pdf}[1][Default]{
  \IfEqCase{#1}{%
      {Default}{\ensuremath{f}}           %
        {DNS}{\ensuremath{f}_{DNS}}           %
        {filter}{\ensuremath{f}_{c}}    %
    }[\PackageError{pdf}{Undefined option to pdf: #1}{}]%
}

\newcommand{\lset}[1][Default]{
    \IfEqCase{#1}{%
        {Default}{\ensuremath{\varphi}}         %
        {DNS}{\ensuremath{\varphi}_{DNS}} 
        {prime}{\ensuremath{\varphi^{\prime}}}      %
        {filter}{\ensuremath{\varphi}_{c}}  %
        {fluctu}{\ensuremath{\tilde{\varphi}}}  %
    }[\PackageError{lset}{Undefined option to lset: #1}{}]%
}%

\newcommand{\meancurv}[1][Default]{
    \IfEqCase{#1}{%
        {Default}{\ensuremath{H}}         %
        {DNS}{H_{DNS}}      %
        {prime}{\ensuremath{H^{\prime}}}      %
        {max}{\ensuremath{H_{\max}}} 			%
        {fluctu}{\ensuremath{\tilde{H}}}  %
        {filter}{\ensuremath{H_{c}}}%
        {var}{\ensuremath{H_{ \lambda}}}%
        {delta}{\ensuremath{\delta H}}%
        {moy}{\ensuremath{\overline{H}}}%
        {K}{\ensuremath{H_{K}}}%
    }[\PackageError{meancurv}{Undefined option to meancurv: #1}{}]%
}%

\newcommand{\gausscurv}[1][Default]{
    \IfEqCase{#1}{%
        {Default}{\ensuremath{G}}         %
        {DNS}{\ensuremath{G_{DNS}}}       %
        {prime}{\ensuremath{G^{\prime}}}      %
        {filter}{\ensuremath{G}_{c}}  %
    }[\PackageError{gausscurv}{Undefined option to gausscurv: #1}{}]%
}%

\NewDocumentCommand{\interfarea}{o o }{%
  \IfNoValueTF{#1}
    {\ensuremath{\Sigma}} %
    {%
        \IfNoValueTF{#2}
          {\IfEqCase{#1}{
        		{Default}{\ensuremath{\Sigma}}        %
        		{DNS}{\ensuremath{\Sigma_{DNS}}}      %
        		{prime}{\ensuremath{\Sigma^{\prime}}}     %
        		{filter}{\ensuremath{\Sigma_{c}}}   %
        		{fluctu}{\ensuremath{\tilde{\Sigma}}}   %
        		{ref}{\ensuremath{\Sigma_{0}}}   %
        		{c}{\ensuremath{\Sigma_{c}}}  %
        		{var}{\ensuremath{\Sigma_{\lambda}}}  %
        		{s}{\ensuremath{\Sigma}}  %
       		{K}{\ensuremath{\Sigma_{K}}}  %
        		{min}{\ensuremath{\Sigma_{\min}}}  %
        		{spot}{\ensuremath{f_{\Sigma}}}  %
        		{ext}{\ensuremath{\Sigma^{\mathrm{ext}}}}%
          }[\ensuremath{\Sigma_{\ite[#1]}}]%
          }%
          {\IfEqCase{#1}{%
        		{fluid}{\ensuremath{\Sigma_{\ite[#1],\ite[#2]}}}%
        		{fluctu}{\IfEqCase{#2}{%
        			{var}{\ensuremath{\tilde{\Sigma}_{\lambda}}}%
        			{delta}{\ensuremath{\delta \tilde{\Sigma}}}  %
        			}[\PackageError{interfarea}{Undefined option with entry fluctu to interfarea: #1}{}]%
        			}%
          	}[\ensuremath{\Sigma_{\ite[#1],\ite[#2]}}]%
          }
    }
}
\NewDocumentCommand{\normal}{o o }{%
  \IfNoValueTF{#1}
    {\ensuremath{\vb{n}}} %
    {%
        \IfNoValueTF{#2}
          {\IfEqCase{#1}{
         {s}{\ensuremath{n}}%
        {Default}{\ensuremath{\bs{n}}}%
        {DNS}{\ensuremath{\vb{n}_{DNS}}}%
        {filter}{\ensuremath{\vb{n}}_{c}}%
        {face}{\ensuremath{\vb{n}}_{\ite[face]}}%
          }[\ensuremath{\vb{n}_{\ite[#1]}}]
          }%
          {\IfEqCase{#1}{
        		{fluid}{\ensuremath{\qty(\vb{n}_{\ite[fluid]})_{\ite[#2]}}}%
          	{var}{\ensuremath{\vb{n}_{\ite[#2], \ite[var]}}}%
          	}[\PackageError{normal}{Undefined option with 2 entries to normal: #1}{}]
          }
    }
}

\NewDocumentCommand{\volfrac}{o o }{%
  \IfNoValueTF{#1}
    {\ensuremath{\ensuremath{\alpha}}} %
    {%
        \IfNoValueTF{#2}
          {\IfEqCase{#1}{
        	{fluid}{\alpha_{\ite[fluid]}}%
        	{var}{\ensuremath{\alpha_{\lambda}}}%
        	{moy}{\ensuremath{\overline{\alpha}}}%
        	{delta}{\ensuremath{\delta \alpha}}%
        	{smel}{\ensuremath{\Delta \alpha}}%
        	{filter}{\alpha_{c}}%
		    {fluctu}{\tilde{\alpha}}%
	        {skt}{\tilde{\alpha}_{k}}%
        	{skpot}{f_{k}}%
        	{s1pot}{f_{1}}%
        	{s2pot}{f_{2}}%
	        {x1}{\ensuremath{\alpha_{x_{1}}}}%
    	    {x2}{\ensuremath{\alpha_{x_{2}}}}%
        	{x3}{\ensuremath{\alpha_{x_{3}}}}%
	        {xi}{\ensuremath{\alpha_{x_{i}}}}%
          }[\ensuremath{\alpha_{\ite[#1]}}]
          }%
          {\IfEqCase{#1}{
        	{moy}{\ensuremath{\overline{\alpha}_{\ite[#2]}}}%
        	{fluid}{\IfEqCase{#2}{
        		{moy}{\ensuremath{\overline{\alpha}_{\ite[#1]}}}%
        		}[\ensuremath{\qty(\alpha_{\ite[fluid]})_{\ite[#2]}}]%
        		}%
          	{var}{\ensuremath{\alpha_{\ite[#2], \ite[var]}}}%
          	}[\PackageError{volfrac}{Undefined option with 2 entries to volfrac: #1}{}]
          }
    }
}
\DeclareRobustCommand{\ball}[1][Default]{
    \IfEqCase{#1}{%
      {Default}{\mathcal{B}}%
      {c}{\mathcal{B}_{c}}%
      {filter}{\mathcal{B}_{c}}%
    }[\PackageError{ball}{Undefined option to ball: #1}{}]%
}%

\NewDocumentCommand{\surface}{o }{%
  \IfNoValueTF{#1}
    {\ensuremath{\mathscr{S}}}%
  {\IfEqCase{#1}{
      {face}{\ensuremath{\mathcal{S}_{\ite[face]}}}%
      {maths}{\ensuremath{\mathscr{S}}}%
      }
      [\ensuremath{\mathscr{S}_{\ite[#1]}}]%
    }
}

\NewDocumentCommand{\spaces}{o o o }{%
  \IfNoValueTF{#1}
    {\ensuremath{\mathbb{R}}}%
    {%
        \IfNoValueTF{#2}
          {\IfEqCase{#1}{
        	{M}{\mathcal{M}}%
        	{TM}{\mathcal{TM}}%
        	{TqM}{\mathcal{T}_{\pathM}\mathcal{M}}%
       	 	{TR}{\mathcal{TR}}%
        	{TR7}{\mathcal{TR}^{7}}%
        	{R7}{\mathbb{R}^{7}}%
        	{R77}{\mathbb{R}^{7\times7}}%
        	{R2}{\mathbb{R}^{2}}%
        	{R3}{\mathbb{R}^{3}}%
        	{R5}{\mathbb{R}^{5}}%
        	{R4}{\mathbb{R}^{4}}%
        	{R}{\mathbb{R}}%
        	{Rp}{\mathbb{R}^{p}}%
        	{Rd}{\mathbb{R}^{d}}%
        	{Rpp}{\mathbb{R}^{p\times p}}%
        	{Rplus}{\mathbb{R}_{+}}%
       	 	{up}{\Omega} %
        	{upelem}{\omega} %
        	{fluid}{\mathcal{N}_{f}}%
        	{fspe}{\mathcal{N}_{s}(\ite[fluid])}%
        	{f1spe}{\mathcal{N}_{s}(1)}%
        	{f2spe}{\mathcal{N}_{s}(2)}%
        	{lfspe}{\mathcal{N}_{s}(\ite[lfluid])}%
        	{spe}{\mathcal{N}_{s}}%
        	{setvar}{\mathcal{O}}%
          	}[\ensuremath{\mathcal{R}^{\ite[#1]}}]
          }%
          {\IfNoValueTF{#3}
            {\IfEqCase{#1}{%
        	{fspe}{\mathcal{N}_{s}(\ite[#2])}%
				{fluid}{\ensuremath{\qty(\vb{n}_{\ite[fluid]})_{\ite[#2]}}}%
	        	{setvar}{\IfEqCase{#2}{
	        		{entropy}{\ensuremath{\mathcal{O}_{\zeta_{\eta}}}}%
              		{natural}{\ensuremath{\mathcal{O}_{\zeta}}}%
              		{naturalk}{\ensuremath{\mathcal{O}_{\zeta_{k}}}}%
              		{pressure}{\ensuremath{\mathcal{O}_{\pi}}}%
              		{pressurek}{\ensuremath{\mathcal{O}_{\pi_{k}}}}%
             		{thermo}{\ensuremath{\mathcal{O}_{\xi}}}%
              		{thermok}{\ensuremath{\mathcal{O}_{\xi_{k}}}}%
              		{thermoext}{\ensuremath{\mathcal{O}_{\Xi_{k}}}}%
              		{entropyvar}{\ensuremath{\mathcal{O}_{\eta_{\lambda}}}}%
					}[\PackageError{spaces}{Undefined option with 2 entries to setvar in spaces: #2}{}]}%
          	}[\PackageError{spaces}{Undefined option with 2 entries to spaces: #1}{}]}%
            {\IfEqCase{#1}{%
	        	{setvar}{\IfEqCase{#2}{
					{natural}{\IfEqCase{#3}{%
            			{fluidiso}{\ensuremath{\mathcal{O}_{\overline{\zeta_{\ite[fluid]}}}}}%
            			}[\ensuremath{\mathcal{O}_{\zeta_{\ite[#3]}}}]%
						}%
					{pressure}{\IfEqCase{#3}{%
            			{fluidiso}{\ensuremath{\mathcal{O}_{\overline{\pi_{\ite[fluid]}}}}}%
            			}[\ensuremath{\mathcal{O}_{\pi_{\ite[#3]}}}]%
						}%
					{thermo}{\IfEqCase{#3}{%
            			{fluidiso}{\ensuremath{\mathcal{O}_{\overline{\xi_{\ite[fluid]}}}}}%
            			}[\ensuremath{\mathcal{O}_{\xi_{\ite[#3]}}}]%
						}%
              		{thermoext}{\ensuremath{\mathcal{O}_{\Xi_{\ite[#3]}}}}%
					}[\PackageError{spaces}{Undefined option with 3 entries to setvar in spaces: #2}{}]}%
              }[\PackageError{speed}{Undefined option with 3 entries to speed: #1}{}]}
          }
    }
}

\newcommand{\spacetx}[1]{%
    \IfEqCase{#1}{%
        {0}{\mathcal{V}}%
        {t}{\mathcal{V}_{t}}%
        {t0}{\mathcal{V}_{t_{0}}}%
        {t1}{\mathcal{V}_{t_{1}}}%
    }[\PackageError{spacetx}{Undefined option to spacetx: #1}{}]%
}%

\newcommand{\capillarity}[1][Default]{
	\IfEqCase{#1}{%
		{Default}{\ensuremath{\gamma}}         %
		{filter}{\ensuremath{\gamma_{c}}}       %
		{fluctu}{\ensuremath{\tilde{\gamma}}}      %
	}[\PackageError{capillarity}{Undefined option to capillarity: #1}{}]%
}%
\NewDocumentCommand{\viscosity}{o }{%
  \IfNoValueTF{#1}
    {\ensuremath{\ensuremath{\mu}}}%
  {\IfEqCase{#1}{
      {add}{\ensuremath{\mu_{add}}}%
      }
      [\ensuremath{\mu_{\ite[#1]}}]%
    }
}

\NewDocumentCommand{\Idmat}{o }{%
  \IfNoValueTF{#1}
    {\ensuremath{\fontmat{I}_{d}}}%
  {\IfEqCase{#1}{
      {7}{\ensuremath{\fontmat{I}_{7}}}%
      {3}{\ensuremath{\fontmat{I}_{3}}}%
      }
      [\ensuremath{\fontmat{I}_{\ite[#1]}}]%
    }
}
\newcommand{\average}[1]{\ensuremath{\left\langle #1 \right\rangle}}

\newcommand{\ddmat}[1]{%
    \IfEqCase{#1}{%
        {1}{D_{1,t}}%
        {2}{D_{2,t}}%
        {k}{D_{k,t}}%
        {kp}{D_{k^{\prime},t}}%
    }[\PackageError{ddmat}{Undefined option to ddmat: #1}{}]%
}%

\NewDocumentCommand{\eigen}{o o }{%
  \IfNoValueTF{#1}
    {\ensuremath{\fontvec{r}}} %
    {%
        \IfNoValueTF{#2}
          {\IfEqCase{#1}{
        		{vec}{\ensuremath{\fontvec{r}}}%
        		{rvec}{\ensuremath{\fontvec{r}}}%
        		{lvec}{\ensuremath{\fontvec{l}}}%
        		{val}{\ensuremath{\lambda}}%
          }[\PackageError{eigen}{Undefined option with 1 entries to eigen: #1}{}]
          }%
          {\IfEqCase{#1}{
        		{rvec}{\ensuremath{\fontvec{r}_{\ite[#2]}}}%
        		{lvec}{\ensuremath{\fontvec{l}_{\ite[#2]}}}%
        		{val}{\ensuremath{\lambda_{\ite[#2]}}}%
          	}[\PackageError{eigen}{Undefined option with 2 entries to eigen: #2}{}]
          }
    }
}

\NewDocumentCommand{\reynolds}{o o }{%
  \IfNoValueTF{#1}
    {\ensuremath{\text{Re}}} %
    {%
        \IfNoValueTF{#2}
          {\IfEqCase{#1}{%
        		{add}{\ensuremath{\text{Re}_{add}}}%
          }[\ensuremath{\text{Re}_{\ite[#1]}}]%
          }%
          {\IfEqCase{#1}{%
        		{add}{\ensuremath{\ensuremath{\text{Re}_{add}}}}%
          	}[\PackageError{reynolds}{Undefined option with 2 entries to reynolds: #2}{}]%
          }
    }
}
\NewDocumentCommand{\weber}{o o }{%
  \IfNoValueTF{#1}
    {\ensuremath{\text{We}}} %
    {%
        \IfNoValueTF{#2}
          {\IfEqCase{#1}{%
        		{add}{\ensuremath{\text{We}_{add}}}%
          }[\ensuremath{\text{We}_{\ite[#1]}}]%
          }%
          {\IfEqCase{#1}{%
        		{add}{\ensuremath{\ensuremath{\text{We}_{add}}}}%
          	}[\PackageError{weber}{Undefined option with 2 entries to weber: #2}{}]%
          }
    }
}

\NewDocumentCommand{\momentumratio}{o o }{%
  \IfNoValueTF{#1}
    {\ensuremath{\text{M}}} %
    {%
        \IfNoValueTF{#2}
          {\IfEqCase{#1}{
        		{add}{\ensuremath{\text{M}_{add}}}%
          }[\ensuremath{\text{M}_{\ite[#1]}}]
          }%
          {\IfEqCase{#1}{
        		{add}{\ensuremath{\ensuremath{\text{M}_{add}}}}%
          	}[\PackageError{momentumratio}{Undefined option with 2 entries to momentumratio: #2}{}]
          }
    }
}

\DeclareCiteCommand{\citejournal}
  {\usebibmacro{prenote}}
  {\usebibmacro{citeindex}%
    \usebibmacro{journal}}
  {\multicitedelim}
  {\usebibmacro{postnote}}

\newcommand{\citeay}[1]{\autocite{#1}} 
\usepackage[pdftex,breaklinks, urlcolor=blue,pdftitle={Work Report},pdfauthor={Pierre CORDESSE},pdfsubject={LaTeX}]{hyperref}
\hypersetup{
  colorlinks   = true,
  citecolor    = blue,
  linkcolor = blue,
}

\title{Initiation of a validation strategy of reduced-order two-fluid flow models using direct numerical simulations in the context of jet atomization}

\shorttitle{Comparative study of jet atomization simulations}

\author{%
    P. Cordesse \and
    A. Murrone \and
    T. Menard \and
    M. Massot
}

\shortauthor{P. Cordesse et~al.}

\makeatletter%
\begin{document}

\setcounter{page}{1}

\maketitle

\begin{abstract}
In industrial applications, developing predictive tools relying on numerical simulations using reduced-order models nourish the need of building a validation strategy. In the context of cryogenic atomization, we propose to build a hierarchy of direct numerical simulation test cases to assess qualitatively and quantitatively diffuse interface models. The present work proposes an initiation of the validation strategy with an air-assisted water atomization using a coaxial injector.
\end{abstract} 
\section{Introduction}
Rocket engines reliability is one of the main priorities given to the European Ariane 6 program. Cryogenic combustion chambers encompass several interacting physical phenomena covering a large spectrum of scales. In particular, the primary atomization plays a crucial role in the way the engine works, thus must be thoroughly studied to understand its impact on high frequencies instabilities. In sub-critical condition, the two-phase flow topologies encountered in the chamber vary from two separated phases at the exit of the nozzle to a polydisperse spray of droplets downstream. In between, liquid structure such as ligaments, rings and deformed droplets detach from the liquid core thus the subscale physics and the topology of the flow in this mixed region is very complex. Predictive numerical simulations are mandatory, at least as a complementary tool to experiments to understand the physics and eventually to conceive new combustion chambers and predict instabilities they may generate.

Since direct numerical simulations of these two-phase flows in realistic configurations are still out of reach, computational power needs being too high, reduced-order models are deployed. However great care must be taken on the choices of these models to combine sound mathematics properties with satisfying representativeness.

Two approaches are found in the literature to build reduced-order models for jet atomization:\\
1. couple two models, a first one for the dispersed flow by employing an element derived from the Kinetic Based Moment Method \citeay{Sibra_2017} which treats polydispersed droplets in size, velocity and temperature, and a second one dealing with the separated phase and the mixed region with either a hierarchy of diffuse interface models \citeay{Drui_JFM_2019} or some level of Large Eddy Simulations (LES) on the interface dynamics \citeay{Herrmann_2013}.\\
2. use a unified model that encompasses any flow topology. Works in this direction are found in \citeay{Drui_JFM_2019} where a unified model accounting for micro-inertia and micro-viscosity associated to bubble pulsation is proposed.

In the present study, we focus on the first approach. We eventually want to perform a numerical simulation of the primary atomization by using a Kinetic-Based-Moment Methods (KBMM) modelling the dispersed flow as in \citeay{Sibra_2017} coupled with a hierarchy of diffuse interface models describing the separated phases and the mixed zone. In order to qualitatively and quantitatively assess the diffuse interface models, we propose a validation strategy based on a hierarchy of direct numerical simulation test cases.  %
\section{Mathematical modelling}

\subsection{Reduced order methods - diffuse interface model}

Among the hierarchy of the diffuse interface models, the Baer and Nunziato model first introduced in \citeay{Baer_Nunziato_1986} has been generalized in \citeay{Saurel_1999} by introducing the interfacial quantities $\pressure[int]$ and $\speed[s][int]$, the interfacial pressure and the interfacial velocity respectively, to model the dynamic of the interface. The system of equation takes the form in one dimension:
\begin{align}\label{sys:BNZ}
\begin{IEEEeqnarraybox}[\IEEEeqnarraystrutmode
\IEEEeqnarraystrutsizeadd{1pt}{1pt}][c]{c}
  \fpdv{t} \CV + \left\lbrace \partial_{\CV} \CF(\CV) + \NCF(\CV) \right\rbrace \partial_{x} \CV = \frac{\source[relax](\CV)}{\bs{\epsilon}} \\ \\
  \text{with } \partial_{\CV} \CF(\CV) = \begin{pmatrix}
        0 & 0 & 0 \\
        0 	& \partial_{\CV[2]} \CF[2](\CV[2]) & 0  \\
    	0	& 0 & \partial_{\CV[1]} \CF[1](\CV[1])
    \end{pmatrix}, \ %
\NCF(\CV) = \begin{pmatrix}
        \speed[s][int] & 0 & 0\\
        \NCFtwo & 0 & 0\\
    	\NCFone & 0 & 0
    \end{pmatrix}
\end{IEEEeqnarraybox}
\end{align}
with $\CV \in \spaces[R7]$, the state vector defined by $\CV = \left( \vol[frac][2], \CV[2], \CV[1] \right)^{t}$, $\CV[fluid] = ( \vol[frac][fluid] \density[fluid], \allowbreak \allowbreak \vol[frac][fluid] \density[fluid] \speed[s][fluid], \allowbreak \vol[frac][fluid] \density[fluid] \energy[tot][fluid] )$, $\CF(\CV) \in \spaces[R7]$ the conservative flux $\CF(\CV) = (0, \CF[2](\CV[2]), \CF[1](\CV[1]))^{t}$ with $\CF_{k}(\CV[fluid]) = (\vol[frac][fluid] \density[fluid] \speed[s][fluid],\allowbreak \vol[frac][fluid] \density[fluid] \speed[s][fluid]^{2}+\vol[frac][fluid] \pressure[fluid],\allowbreak \vol[frac][fluid] ( \density[fluid] \energy[tot][fluid]+\pressure[fluid])\speed[s][fluid] )^{t}$, $\NCF(\CF) \in \spaces{R}^{7 \times 7}$ the non-conservative terms with $\NCFtwo(\CV) = - \NCFone(\CV) = (0,\allowbreak  -\pressure[int], \allowbreak -\pressure[int] \speed[s][int])^{t}$, $\vol[frac][fluid]$ the volume fraction of phase $k \in \left\lbrace 1,2 \right\rbrace$, $\density[fluid]$ the partial density, $\speed[s][fluid]$ the phase velocity, $\pressure[fluid]$ the phase pressure, $\energy[tot][fluid]=\energy[fluid] + \allowbreak \speed[s][fluid]^{2}/2$ the total energy per unit of mass, $\energy[fluid]$ the internal energy, $\source[relax](\CV)/\bs{\epsilon} \in\spaces[R7]$ the relaxation terms detailed in the sequel.

Mathematical properties of System~\eqref{sys:BNZ} have been widely examined in the literature \citeay{Embid_Baer_1992, Coquel_2002, Gallouet_2004}: it is a first-order non-linear system of partial differential equations which include conservative and non-conservative terms; the system is conditionally hyperbolic and admits seven eigenvalues,
\begin{align}
	\left\lbrace \speed[s][int], \left\lbrace \speed[s][fluid] , \speed[s][fluid] \pm \speed[sound][fluid] \right\rbrace_{k=2,1} \right\rbrace
\end{align}
with $\speed[sound][fluid]$ the phase sound speed defined by $\speed[sound][fluid]^{2} = \left( \frac{\partial \pressure[fluid]}{\partial \density[fluid]} \right)_{\entropy[fluid]}$.

The interfacial terms need a closure and the thermodynamics must be postulated. A recent work \citeay{Cordesse_2019_CMS} has extended the theory of first- and second-order non-linear conservative systems \citeay{Godunov_1961, Kawashima_1988} to non-conservative systems, and has applied it to the Baer-Nunziato model to obtain a compatible thermodynamics and model closure. All the existing closures in the literature are recovered.

The interfacial velocity $\speed[s][int]$ is commonly chosen in order to be linearly-degenerated which restricts its definition to $\speed[s][int] = \beta \speed[s][1] + (1-\beta) \speed[s][2]$ with $\beta \in \left\lbrace 0,1, \vol[frac][1] \density[1]/\density \right\rbrace$ \citeay{Coquel_2002}.

For the present work, the phases are assumed immiscible such that the postulated entropy $\MES$ takes the form:
\begin{align}\label{eq:mixture_entropy_immiscible}
	\MES = - \sum_{k=1,2} \vol[frac][fluid] \density[fluid] \entropy[fluid]
\end{align}
with $\entropy[fluid]=\entropy[fluid](\density[fluid],\pressure[fluid])$ the phase entropy following a two-parameter equation of state.

The relaxation of the two phases towards an equilibrium state is described by the sources terms $\Srelax/\epsilon$ in System~\eqref{sys:BNZ}. Only mechanical, and hydrodynamic relaxations are accounted for in the present study and therefore $\Srelax/\epsilon$ decomposes into:
\renewcommand\arraystretch{1}
\begin{align}\label{eq:relax_term_7eq}
	\frac{\Srelax}{\epsilon} = \frac{\source[relax][speed]}{\epsilon_{u}} + \frac{\source[relax][pressure]}{\epsilon_{p}}, \text{ with } \frac{\source[relax][speed]}{\epsilon_{u}} =  \left( 0 , \frac{\source[relax][speed]_{2}}{\epsilon_{u}}, \frac{\source[relax][speed][1]}{\epsilon_{u}} \right)^{t} \text{ and } \frac{\source[relax][pressure]}{\epsilon_{p}} = \left( \frac{\pressure[2] - \pressure[1]}{\epsilon_{p}}, \frac{\source[relax][pressure][2]}{\epsilon_{p}}, \frac{\source[relax][pressure][1]}{\epsilon_{p}} \right)^{t}
\end{align}
where $\source[relax][speed][2] = -\source[relax][speed][1] =  \left( 0 , \speed[s][2] - \speed[s][1], \speed[s][int] (\speed[s][2] - \speed[s][1])\right)$, $ \source[relax][pressure][2] = -\source[relax][pressure][1] = \left( 0 , 0 , \pressure[int] (\pressure[2] - \pressure[1]) \right)$ and $\epsilon_{p}$ (resp. $\epsilon_{u}$) is the characteristic time of the mechanical relaxation (resp. hydrodynamical relaxation). When relaxing also the temperatures, one obtains the compressible multi-species Navier-Stokes equations, called also \textit{four equation model}. The Baer-Nunziato model introduced herein before is thus called the \textit{seven equation model}.
\subsection{Incompressible Navier-Stokes equations and VOF for DNS}
In order to describe two-phase flows, the incompressible Navier-Stokes equations are introduced in System~\eqref{eq:navierstokes_diphas}
\begin{eqnarray}
\begin{cases} 
\div{\speed} =0 \\
\fpdv{t} \speed  = - \left( \speed \leftdiv \right)\speed  +\dfrac{1}{\density} \left( -\grad{\pressure} +  \div{\left(2 \mu \Sstrain \right)}  +\textbf{F}   \right)
\label{eq:navierstokes_diphas}
\end{cases}
\end{eqnarray} 
where $\speed \in \spaces{R3}$ is the velocity field, $\mu$ the dynamic viscosity, $\density$ the mixture density, $\pressure$ the equilibrium pressure, $\Sstrain$ the strain tensor defined by $\Sstrain = \frac{1}{2}\left( \nabla(\speed) + \nabla(\speed^{T})\right)$, $\textbf{F}=\textbf{F}_V+\textbf{F}_{ST}$ represents the body force and the surface tension force, $\textbf{F}_{ST}=\capillarity \kappa \delta_I \bs{\vec{n}}$, $\capillarity$ is the surface tension, $\kappa$ the curvature at the interface, $\delta_I$ the Dirac function and $\bs{\vec{n}}$ the interface outwards normal vector.

Then, the interface is implicitly derived by the zero of a Level Set of a scalar function $\phi$ which stands for the signed distance to the interface ($\vert\grad{\phi}\vert=1$, $\phi>0$ in the liquid and $\phi<0$ in the gas). Geometrical properties of the interface, normal and curvature, are easily deduced, $\bs{\vec{n}} = \grad{\phi}/ \vert \grad{\phi} \vert$ and $\kappa= \div{\bs{\vec{n}}}$. The interface motion is captured by the transport of the Level Set function:
\begin{equation}\label{eq:TLS}
\fpdv{t} \phi + \speed \leftdiv \phi=0
\end{equation}
However, the property of distance function is not generally conserved after the resolution of Equation~\eqref{eq:TLS}, an additional equation have to be solved \cite{sussmanredist}. As a consequence of solving both equations, the conservation of mass is not guaranteed. To solve this problem, the volume fraction of liquid (VOF - Volume Of Fluid) $\vol[frac][l]$ in a volume $\Omega$ is introduced in terms of the Level Set function $\phi(\bs{x})$ as:
\begin{align}
\vol[frac][l]= \frac{1}{\Omega} \int_{\Omega} H(\phi(\bs{x})) d\Omega
\end{align}
and the transport of this quantity in the control volume $\Omega$ is given by :
\begin{equation} \label{eq:TVOF}
\fpdv{t} \vol[frac][l] + \speed \leftdiv{\vol[frac][l]} = 0
\end{equation}
The benefits of a VOF formulation coupled with a Level Set function are to conserve mass and to have access to geometrical properties of the interface.
\section{Numerical methods}

\subsection{Diffuse interface model}
The numerical methods employed to solve System~\eqref{sys:BNZ} are implemented in the multiphysics computational fluid dynamics software CEDRE \citeay{Gaillard_2016} working on general unstructured meshes and organized as a set of solver \citeay{Gaillard_2016}. The solver \textit{SEQUOIA} is in charge of the diffuse interface model.

A Strang splitting technique is applied on a multi-slope HLLC with hybrid limiter solver \citeay{Furfaro_Saurel_2015, Letouze_2014} to achieve a time-space second-order accuracy on the discretized equations. The issue of the non-conservative terms encountered when solving System~\eqref{sys:BNZ} is tackled in \citeay{Furfaro_Saurel_2015} by assuming $(1)$ the interfacial terms $\pressure[int]$ and $\speed[s][int]$ to be local constants in the Riemann problem, $(2)$ the volume fraction to vary only across the interfacial contact discontinuity $\speed[s][int]$. As a result, the non conservative terms in System~\eqref{sys:BNZ} vanish, $\speed[s][int]$ and $\pressure[int]$ are determined locally by Discrete Equation Method (DEM) \citeay{Saurel_Gavrilyuk_2003} at each time step and stay constant during the update. Thus, the phases are decoupled and the System~\eqref{sys:BNZ} splits into two conservative sub-systems onto which we apply the multi-slope HLLC with hybrid limiter solver.

Depending on the application, the relaxations are assumed either instantaneous or finite in time. In our case, it is reasonable to assume an instantaneous pressure relaxation but to consider a finite velocity relaxation as in the context of an assisted air-water coaxial injector, the interface dynamic is mainly driven by the shear stress induced by a high velocity gradient at the injection.

To obtain the relaxed pressure, one needs to solve the ODE,
\begin{align}
	\partial_{t} \CV = \frac{ \source[relax][pressure](\CV)}{ \epsilon_{p}},
\end{align}
with $\epsilon_{p} \rightarrow \infty$ which infers $\speed[fluid]$ remains constant. The problem reduces to apply an iterative procedure as a Newton method to solve a second order equation. Details of the equation can be found in \citeay{Furfaro_Saurel_2015}. As for the velocities, the following ODE is solved
\begin{align}
	\partial_{t} \speed[diff] - \frac{ A^{o} \speed[diff] }{ \epsilon_{u}} = 0,
\end{align}
where $\speed[diff]$ is the slip velocity $\speed[diff] = \speed[2]-\speed[1]$ and superscript ${}^{o}$ denotes the state before relaxation. A first numerical approach is to fix a remaining slip velocity ratio target at each computational time step $\Delta t$. It defines the characteristic relaxing time:
\begin{align}
	\frac{\epsilon_{u}}{A^{o}} = \ln{(X)} \Delta t \ \text{with } X = \frac{\speed[diff](\Delta t)}{\speed[diff]^{o}} \text{ and } A^{o} = \frac{\vol[frac][1]^{o} \density[1]^{o} + \vol[frac][2]^{o} \density[2]^{o}}{\vol[frac][1]^{o} \density[1]^{o}\vol[frac][2]^{o} \density[2]^{o}}
\end{align}
An instantaneous velocity relaxation is in pratice also possible and manipulating the ODE leads to a unique relaxed velocity, which is the mass weighted average of the two velocities before relaxing.

\subsection{DNS numerical methods}
As mentioned by Rudman \cite{rudman}, it is recommended to solve Navier-Stokes equations in their conservative form to ensure consistance between mass fluxes (equivalent to VOF fluxes) and momentum fluxes. Since a staggered grid is adopted, VOF and velocity do not have the same control volume and it is difficult to obtain this consistency. This is the reason why Rudman introduces a grid two times smaller in each direction for VOF transport. In \cite{Vaudor_2017}, Vaudor developed a method to avoid this finest grid. This method is used here and allows to reduce computation time.

The Navier-Stokes equations are implemented in the code ARCHER \cite{Vaudor_2017}. A projection method is employed and physical properties (viscosity and density) are expressed in term of both VOF and Level Set. The temporal integration is performed through a second-order Runge-Kutta scheme. The discretization of convective term is achieved with WENO 5 scheme \cite{efficient_weno}. For the viscosity term, we retain the method presented by Sussman \cite{sussman}. The Ghost-Fluid \cite{Fedkiw} method is employed to take into account surface tension force $F_{ST}$, treated as a pressure jump.

To compute correctly mass fluxes at the interface (consequently momentum fluxes) and consistence between both flow and interface solvers, VOF and Level set transport (Equations~\eqref{eq:TLS} and ~\eqref{eq:TVOF}) are performed with a CLSVOF algorithm \cite{sussman_clsvof} at each step of Runge Kutta scheme. %
\section{Results and discussions}
As part of the validation strategy, the hierarchy of direct numerical simulation test cases starts with an air-assisted water atomization with a coaxial injector and results are presented hereafter. 

The injector at Geophisic and industrial flow laboratory (LEGI) has been the subject of several experiments \citeay{Rehab_1997, Lasheras_2000, Marmottant_2002, Delon_2018} covering a large range of flow conditions and thus offers experimental data to assess numerical simulations. The present simulations reproduce the experiments conducted by \citeay{Delon_2013}, whose results of interests for our study are presented in \citeay{Vaudor_2017} along with the numerical results obtained with ARCHER.

\subsection{Description of the LEGI experiment}

\begin{figure}[h]
\centering
\begin{minipage}[h!]{0.33\linewidth}%
\footnotesize
\begin{overpic}[width=1.0\textwidth]{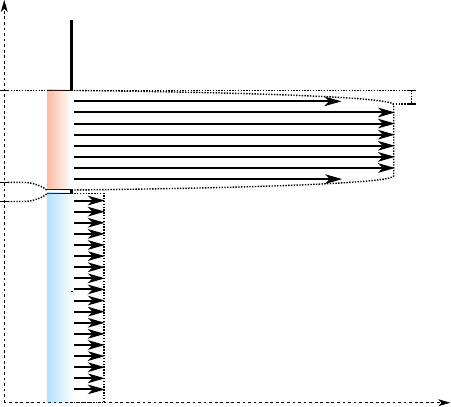}
         \put(30,25){$\speed[S][l]$}
         \put(91,55){$\speed[S][g]$}
         \put(95,67){$\delta$}
         \put(-10,68){$\radius[g]$}
         \put(-10,41){$\radius[l]$}
         \put(-19,50){$\radius[l]\text{+}e$}
         \put(0,93){$r$}
         \put(102,0){$x$}
       \end{overpic}
\end{minipage}%
\hfill
\begin{minipage}[h!]{0.65\linewidth}%
\begin{align}
\speed[S][l] = 0.26 \, m.s^{-1}
\end{align}
\begin{align}\label{eq:gas_velocity}
\begin{IEEEeqnarraybox}[\IEEEeqnarraystrutmode
\IEEEeqnarraystrutsizeadd{1pt}{1pt}][c]{cl}
\speed[S][g] &= \begin{cases} 
      \dfrac{r-\radius[l]-e}{\delta}\speed[S][g]^{max} & \radius[l]+e \leq r \text{ and} \\
                             &  \ r \leq \radius[l]+e+\delta  \\
      \speed[S][g]^{max} & \radius[l]+e+\delta \leq r \\
              & \text{and} \ r \leq \radius[g]-\delta \\
       \left(1-\dfrac{r-(\radius[g]-\delta)}{\delta}\right)\speed[S][g]^{max} & \radius[g]-\delta \leq r \\
                                      & \text{and} \ r \leq \radius[g]
   \end{cases}\\
   \text{ with } & \delta = 0.2 \, mm \text{ and } \speed[S][g]^{max}=25 \, m.s^{-1}
\end{IEEEeqnarraybox}
\end{align}
\end{minipage}
\caption{Injector schematic and velocity profiles.}\label{fig:injector_geom}
\end{figure}
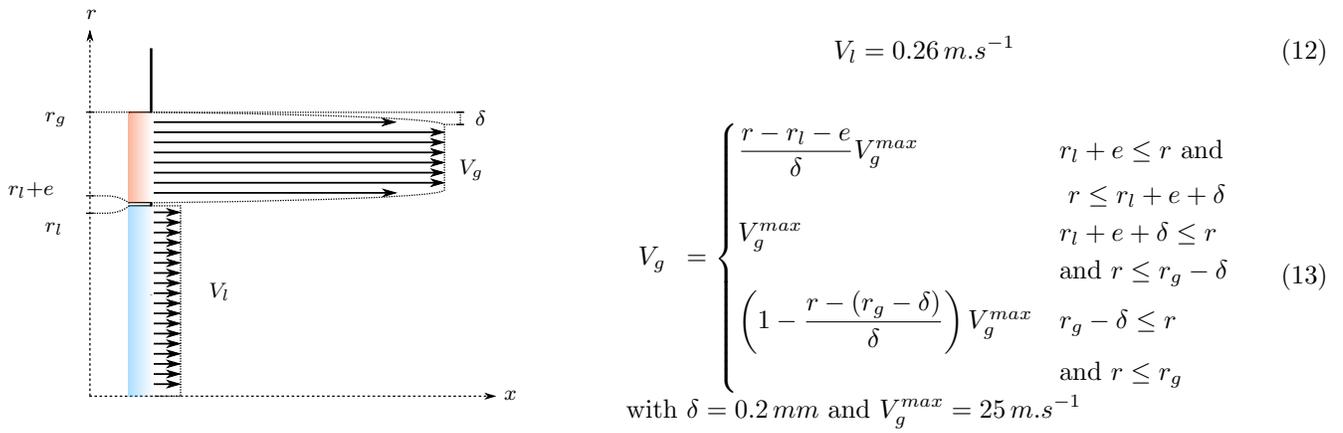

Figure~\ref{fig:injector_geom} renders the simulation configuration which is identical to the experiment of \citeay{Delon_2013} and indicates the velocity profiles at the injector outlets measured experimentally. 

The inside diameter measures $\diam[l] =7,6 \, mm$, the outside diameter $\diam[g] = 11,4 \, mm$, the lip length $e = 0.2 \, mm$. The gas velocity profile $\speed[S][g]$ given in Equation~\eqref{eq:gas_velocity} models the boundary layer measured experimentally with $\delta$ the boundary layer thickness and $\speed[S][g]^{max}$ the maximum gas velocity.

The fluid properties, type, density $\density$, capillarity coefficient $\capillarity$ and viscosity coefficient $\viscosity$ of each fluid are summarized in Table~\ref{table:physical_parameters}.
\begin{table}[h]
\renewcommand{\arraystretch}{1,5}
\centering
    \begin{tabular}{c|c|c|c|c}
      & Phase   & $\density$ $(kg.m^{-3})$       & $\capillarity$ $(N.m^{-1})$ & $\viscosity$ $(1e^{-5}Pa.s)$ \\ \hline
        Liquid $(l)$  & $H_{2}O$  & $1000$  & $0.0072$  & $1002$ \\ \hline
        Gas   $(g)$ & $Air$   & $1.226$   & $0.0072$  & $17.8$
    \end{tabular}
    \caption{Physical properties of water and air.}
    \label{table:physical_parameters}
\end{table}

\begin{figure}[h]
    \centering
      \begin{overpic}[width=0.7\textwidth ]{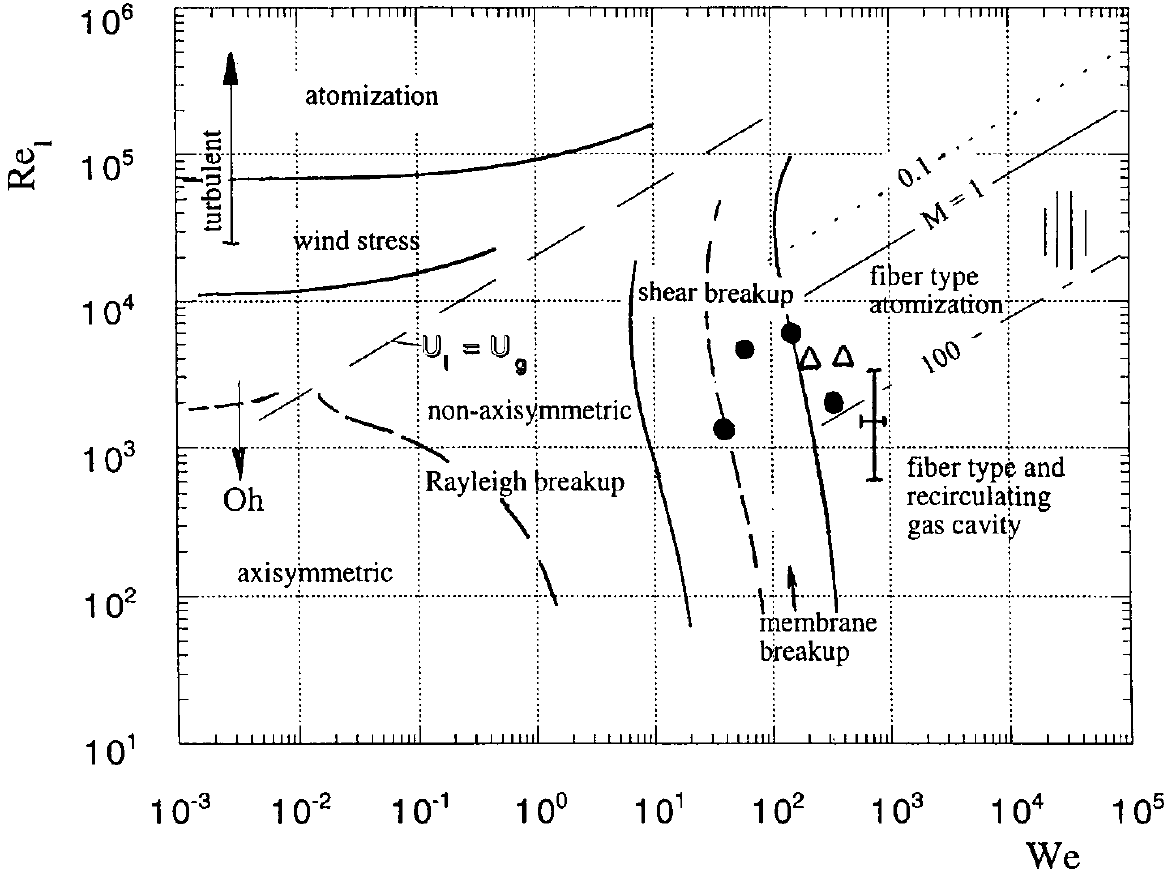} %
         \put(64,37){\color{red}\circle*{2.5}}
      \end{overpic}
    \caption[{Break-up regimes in the parameter spaces $\reynolds[l] - \weber$}]{Break-up regimes in the parameter spaces $\reynolds[l] - \weber$ \citeay{Lasheras_2000} - LEGI \protect\fillcircle{red}}
    \label{fig:regime_atom}
\end{figure}

Let us define and compute the following flow parameters. $\reynolds[l]$ is the liquid Reynolds number, $\reynolds[g]$ is the gas Reynolds number, $\momentumratio$ the momentum flux ratio, $\weber[l]$ the liquid Weber Number, $\weber[g]$ the gas Weber Number and $\weber$ the aerodynamic Weber number. They are defined as follows
\begin{subequations}
\begin{align}
  \reynolds[l] = \frac{\density[L] \speed[S][L] \diam[l]}{\viscosity_{L}} = 1972,
   \, 
  \reynolds[g] = \frac{\density[G] \speed[S][G] (\diam[g]-\diam[l]-2e)}{\viscosity_{G}}=5854,
   \, 
   \momentumratio = \frac{\density[G] \speed[S][G]^{2}}{\density[L] \speed[S][L]^{2}} = 11,
\end{align}
\begin{align}
  \weber[l] = \frac{\density[L] \speed[S][L]^{2} \diam[l]}{\capillarity} = 7,
  \, 
  \weber[g] = \frac{\density[G] \speed[S][G]^{2} (\diam[g]-\diam[l]-2e)}{\capillarity} = 36,
  \, 
  \weber = \frac{\density[G] \speed[S][G]^{2}  \diam[l]}{\capillarity} = 81.
\end{align}
\end{subequations}
The Weber numbers are small inferring the surface tension should not play a crucial role in the dynamics of the flow. For the present configuration, there exists a break-up regime map established by \citeay{Lasheras_2000}, reported in Figure~\ref{fig:regime_atom}.
The red dot of coordinates $(\reynolds[l]=1972, \weber = 81)$ locates the investigated flow in the shear breakup zone in Figure~\ref{fig:regime_atom}. We should not expect atomization of the liquid jet and therefore the atomization is not activated in CEDRE.

\subsection{Numerical set-up}

The ARCHER simulations are performed on a Cartesian mesh $1024 \, \times \, 512  \, \times \,  512$ with a cell size equal to $\Delta x = 6.68 \, 10^{-5}$, so a total of $806$ millions of faces whereas CEDRE mesh is composed of $278\, 978$ tetrahedral cells. The mesh size is $563 \,116$ faces.

\begin{figure}[h]
\centering
\subfloat[CEDRE mesh]{%
       \centering
  \includegraphics[width=0.5\textwidth]{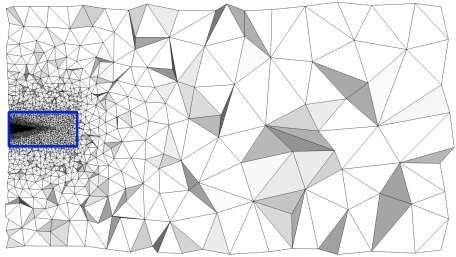} %
  \label{subfig-1:cedre_mesh}
}
\subfloat[Refined mesh box of the dimension of the DNS geometry]{%
       \centering
  \includegraphics[width=0.5\textwidth]{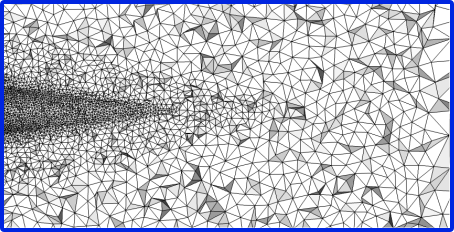} %
  \label{subfig-2:cedre_mesh}
}
\caption{CEDRE mesh of the configuration}
\label{fig:cedre_mesh}
\end{figure}

To be comparable to the DNS, two difficulties must be tackled. Firstly, the DNS uses an incompressible solver meaning the acoustic is not solved and thus does not interact with the liquid jet and the density remains constant. Secondly, the boundary conditions imposed on the wall of the DNS box do not let any flow backwards such that no wall effect acts upon the liquid core. Therefore, to eliminate reflected acoustic waves and wall effect on the liquid jet, we have designed an outer box with a coarsening mesh as shown in Figure~\ref{fig:cedre_mesh}.
The smallest cell is located at the lip of the injector and measures $\Delta x = 2.0 \,10^{-4} \, m$. As for the thermodynamics, CEDRE uses two Stiffened-Gas equations of state and thus the temperature of the phases has been modified to obtain the same initial pressure and density conditions as in Table~\ref{table:physical_parameters}.

The simulation information are summarized in Table~\eqref{table:CPU_info}.
\begin{table}[h]
\renewcommand{\arraystretch}{1,5}
\centering
    \caption{Simulation costs}
    \label{table:CPU_info}
    \begin{tabular}{c|c|c|c}
            &  Simu Time $(s)$    & $N_{proc}$  & Total CPU $(h)$\\ \hline
        CEDRE   & 0.160         & $420$     & $2.14 \, 10^{4}$   \\ \hline
        ARCHER    & 0.500         & $8192$    & $10 \, 10^{6}$
    \end{tabular}
\end{table}
Remarkably, when considering the same simulation time, the total CPU is $150$ greater for ARCHER than for CEDRE.
\subsubsection{Qualitative comparison}
\begin{figure}[h]
\centering
\includegraphics[width=1.0\textwidth]{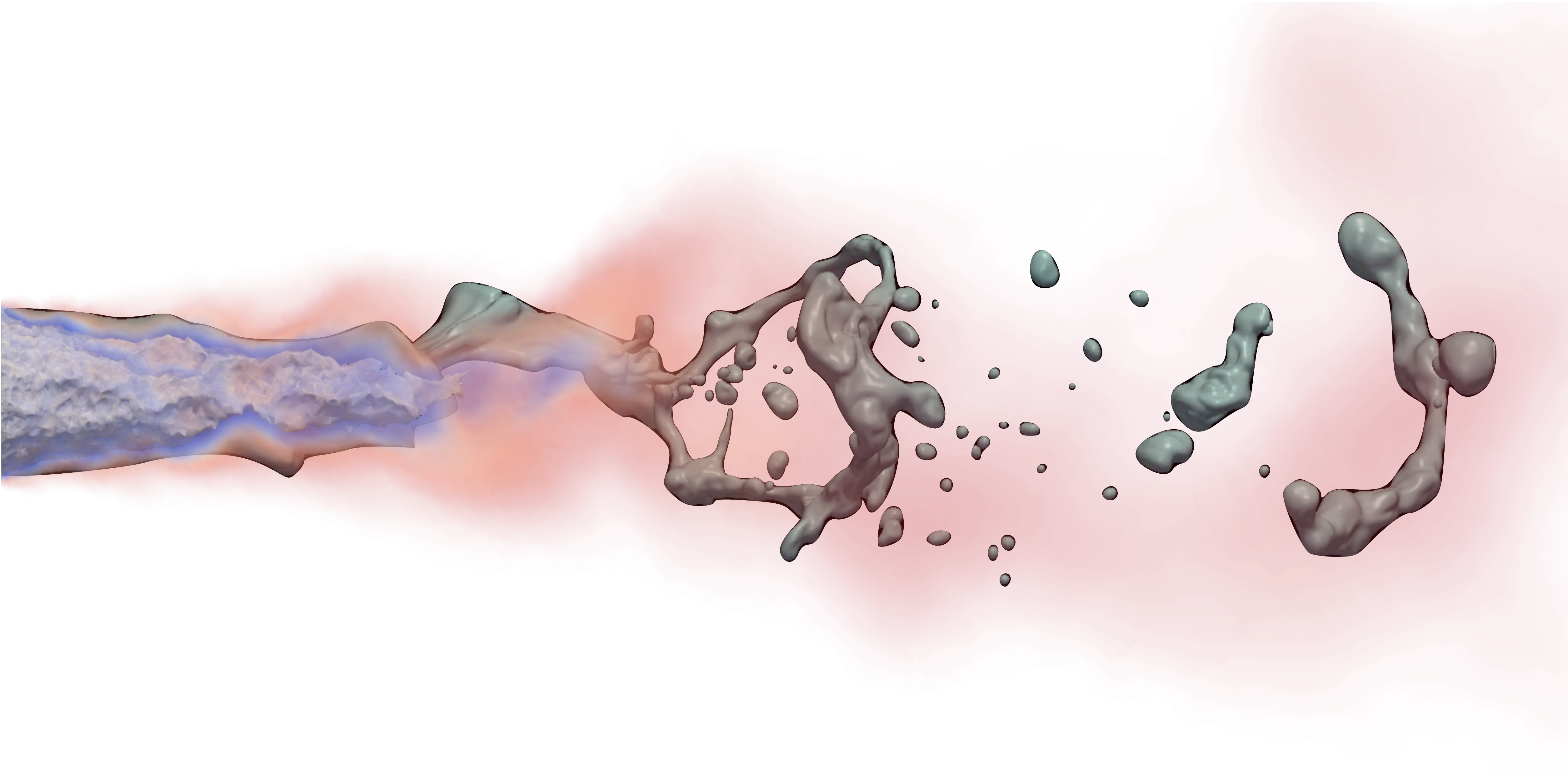} %
\caption{Instantaneous liquid core comparison. CEDRE: volume rendering of the liquid volume fraction, $\vol[frac][l]$ high \protect \includegraphics[height=0.7em, width=5em]{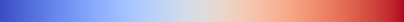} $\ $ low, grey isosurface $\vol[frac][l] = 0.99$. ARCHER: Grey isovolume of liquid volume fraction $\vol[frac][l] = 1$ }
\label{fig:inst_liquid_core}
\end{figure}

We propose to compare the liquid jet obtained with ARCHER and CEDRE at given simulation time. The Level Set function of the DNS permits an exact reconstruction of the interface whereas for the diffuse interface model, the interface lays in the region where the volume fraction varies from $\vol[frac] \approx 0$ to $\vol[frac] \approx 1$. Consequently, in Figure~\ref{fig:inst_liquid_core} we have superimposed the solved interface of the DNS to a volume rendering of the liquid volume fraction and a single liquid volume fraction isosurface.
We distinguish two regions: close to the injector, on the first half of the DNS box, the diffuse interface model is able to match accordingly to the DNS results. In this region, the mesh used by CEDRE prevents the interface to diffuse too much for the diffuse interface model and the interface of the DNS lays in the volume rendering of the liquid volume fraction of CEDRE. On the second half of the DNS box, where the DNS shows complex liquid structures such as rings, droplets and ligaments, the diffuse interface model is not able to capture these effects since the mesh is coarsened and the volume fraction alone is not sufficient to describe such complex interface dynamics. Nevertheless, the \textit{seven equation model} diffusion accords well with the DNS, the red cloud which corresponds to a low liquid volume fraction is limited to the zones where liquid elements of the DNS exists. 

This comparison gives an interesting interpretation of the diffuse interface models. The volume fraction is not enough to reconstruct the whole dynamic of the interface but attests the presence or the absence of liquid.
\subsubsection{Quantitative comparison}
\begin{figure}[!h]\centering
\resizebox{!}{5cm}{%
\begingroup
  \makeatletter
  \providecommand\color[2][]{%
    \GenericError{(gnuplot) \space\space\space\@spaces}{%
      Package color not loaded in conjunction with
      terminal option `colourtext'%
    }{See the gnuplot documentation for explanation.%
    }{Either use 'blacktext' in gnuplot or load the package
      color.sty in LaTeX.}%
    \renewcommand\color[2][]{}%
  }%
  \providecommand\includegraphics[2][]{%
    \GenericError{(gnuplot) \space\space\space\@spaces}{%
      Package graphicx or graphics not loaded%
    }{See the gnuplot documentation for explanation.%
    }{The gnuplot epslatex terminal needs graphicx.sty or graphics.sty.}%
    \renewcommand\includegraphics[2][]{}%
  }%
  \providecommand\rotatebox[2]{#2}%
  \@ifundefined{ifGPcolor}{%
    \newif\ifGPcolor
    \GPcolortrue
  }{}%
  \@ifundefined{ifGPblacktext}{%
    \newif\ifGPblacktext
    \GPblacktextfalse
  }{}%
  \let\gplgaddtomacro\g@addto@macro
  \gdef\gplbacktext{}%
  \gdef\gplfronttext{}%
  \makeatother
  \ifGPblacktext
    \def\colorrgb#1{}%
    \def\colorgray#1{}%
  \else
    \ifGPcolor
      \def\colorrgb#1{\color[rgb]{#1}}%
      \def\colorgray#1{\color[gray]{#1}}%
      \expandafter\def\csname LTw\endcsname{\color{white}}%
      \expandafter\def\csname LTb\endcsname{\color{black}}%
      \expandafter\def\csname LTa\endcsname{\color{black}}%
      \expandafter\def\csname LT0\endcsname{\color[rgb]{1,0,0}}%
      \expandafter\def\csname LT1\endcsname{\color[rgb]{0,1,0}}%
      \expandafter\def\csname LT2\endcsname{\color[rgb]{0,0,1}}%
      \expandafter\def\csname LT3\endcsname{\color[rgb]{1,0,1}}%
      \expandafter\def\csname LT4\endcsname{\color[rgb]{0,1,1}}%
      \expandafter\def\csname LT5\endcsname{\color[rgb]{1,1,0}}%
      \expandafter\def\csname LT6\endcsname{\color[rgb]{0,0,0}}%
      \expandafter\def\csname LT7\endcsname{\color[rgb]{1,0.3,0}}%
      \expandafter\def\csname LT8\endcsname{\color[rgb]{0.5,0.5,0.5}}%
    \else
      \def\colorrgb#1{\color{black}}%
      \def\colorgray#1{\color[gray]{#1}}%
      \expandafter\def\csname LTw\endcsname{\color{white}}%
      \expandafter\def\csname LTb\endcsname{\color{black}}%
      \expandafter\def\csname LTa\endcsname{\color{black}}%
      \expandafter\def\csname LT0\endcsname{\color{black}}%
      \expandafter\def\csname LT1\endcsname{\color{black}}%
      \expandafter\def\csname LT2\endcsname{\color{black}}%
      \expandafter\def\csname LT3\endcsname{\color{black}}%
      \expandafter\def\csname LT4\endcsname{\color{black}}%
      \expandafter\def\csname LT5\endcsname{\color{black}}%
      \expandafter\def\csname LT6\endcsname{\color{black}}%
      \expandafter\def\csname LT7\endcsname{\color{black}}%
      \expandafter\def\csname LT8\endcsname{\color{black}}%
    \fi
  \fi
    \setlength{\unitlength}{0.0500bp}%
    \ifx\gptboxheight\undefined%
      \newlength{\gptboxheight}%
      \newlength{\gptboxwidth}%
      \newsavebox{\gptboxtext}%
    \fi%
    \setlength{\fboxrule}{0.5pt}%
    \setlength{\fboxsep}{1pt}%
\begin{picture}(5102.00,2834.00)%
    \gplgaddtomacro\gplbacktext{%
      \csname LTb\endcsname%
      \put(0,550){\makebox(0,0)[r]{\strut{}\small 0}}%
      \put(0,894){\makebox(0,0)[r]{\strut{}\small 5}}%
      \put(0,1238){\makebox(0,0)[r]{\strut{}\small 10}}%
      \put(0,1582){\makebox(0,0)[r]{\strut{}\small 15}}%
      \put(0,1925){\makebox(0,0)[r]{\strut{}\small 20}}%
      \put(0,2269){\makebox(0,0)[r]{\strut{}\small 25}}%
      \put(0,2613){\makebox(0,0)[r]{\strut{}\small 30}}%
      \put(132,330){\makebox(0,0){\strut{}\small 20}}%
      \put(737,330){\makebox(0,0){\strut{}\small 40}}%
      \put(1341,330){\makebox(0,0){\strut{}\small 60}}%
      \put(1946,330){\makebox(0,0){\strut{}\small 80}}%
      \put(2551,330){\makebox(0,0){\strut{}\small 100}}%
      \put(3155,330){\makebox(0,0){\strut{}\small 120}}%
      \put(3760,330){\makebox(0,0){\strut{}\small 140}}%
      \put(4364,330){\makebox(0,0){\strut{}\small 160}}%
      \put(4969,330){\makebox(0,0){\strut{}\small 180}}%
    }%
    \gplgaddtomacro\gplfronttext{%
      \csname LTb\endcsname%
      \put(-374,1581){\rotatebox{-270}{\makebox(0,0){\strut{}$\text{\small $\lenLC$ $[$mm$]$}$}}}%
      \put(2550,110){\makebox(0,0){\strut{}\small $t$ $[$ms$]$}}%
    }%
    \gplbacktext
    \put(0,0){\includegraphics{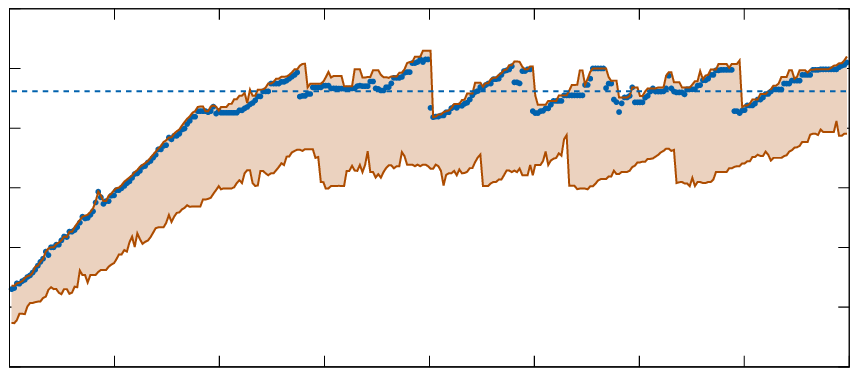}}%
    \gplfronttext
  \end{picture}%
\endgroup
}
\caption{Instantaneous liquid jet length $\lenLC$ over time at isovalue $\vol[frac][2] \in \left[0.91, 1-2\epsilon\right]$, \fillcircle{Blueone} isovalue $\vol[frac][2]=0.99$, \lineplain{Redone} $\vol[frac][2]=1-2\epsilon$ (bottom) and $\vol[frac][2]=0.91$ (top), \linedot{Blueone} $\left\langle \lenLC(\vol[frac][2]=0.99) \right\rangle$ for $t > 60 \, ms$.}
\label{plot:liq_jet_len}
\end{figure}

The liquid core is defined as the region of the liquid jet that is always occupied by liquid. To obtain it, a time-averaging of the liquid volume fraction is needed. To compare the liquid core obtained with CEDRE, we need to eliminate the transient phase during which the liquid jet is not established. Figure~\ref{plot:liq_jet_len} plots the instantaneous liquid jet length $\lenLC$ over time for several $\vol[frac][l]$.

\begin{figure}[h]
    \centering
      \begin{overpic}[width=1.0\textwidth ]{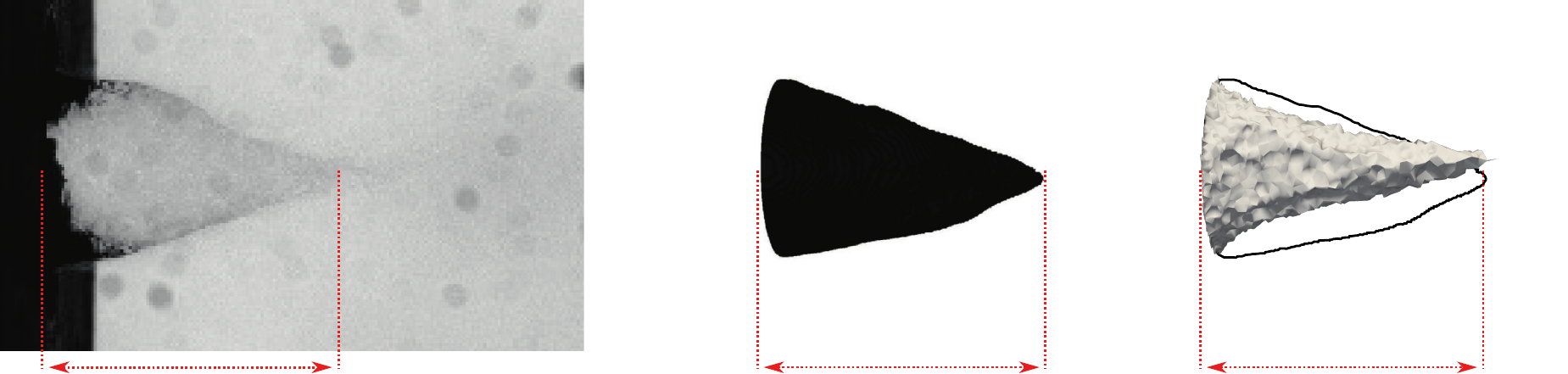} %
         \put(6,-3){$\text{Expe }\lenLC=12.1 \, mm$}
         \put(50,-3){$\text{DNS }\lenLC=12 \, mm$}
         \put(78,-3){$\text{CEDRE }\lenLC=11.8 \, mm$}
      \end{overpic}
      \vspace{1em}
    \caption{Liquid core comparison - from left to right: experiment, DNS, CEDRE and DNS. CEDRE liquid core is the isovolume at $\average{\vol[frac][l]}=1-2 \epsilon$, with $\average{\vol[frac][l]}$ the time averaged liquid volume fraction and $\epsilon=1e{-}6$ residual volume fraction.}
    \label{fig:liquid_core_comparison}
\end{figure}

For $\time > 60 \, ms$ the liquid jet length starts to oscillate around a time averaged value which equals $\left\langle \lenLC(\vol[frac][2]=0.99 \right\rangle = 23 \, mm$. We then computed the time average liquid volume fraction between $\time[sim] \in \left[ 60, 160 \right] \, ms$ to obtain the liquid core shown in Figure~\ref{fig:liquid_core_comparison}.
The liquid core of the diffuse interface model was identified as the isovolume of $\vol[frac][l]=1-2\epsilon$, where $\epsilon=1e-6$ is the residual volume fraction. The reason for not choosing $\vol[frac][l]=1-\epsilon$ holds to the fact that due to numerical diffusion in a unstructured mesh and mesh interpolation, the volume fraction in single phase region does not stay at the initial value $\vol[frac][fluid]=1-\epsilon$. 

Experimental, DNS and CEDRE values show the same trend.

We also could have compared the angle of the spray, but for the diffuse interface model, the result depends highly on the threshold chosen for the liquid volume fraction.
\section{Conclusion}
In the present contribution, we started a validation process of reduced-order models in the context of cryogenic propulsion using direct numerical simulations.

The test case proposed here is an air-assisted water atomization using a coaxial injector which also provides experimental results from the LEGI test bench. The comparison showed good agreements and important CPU gains between the \textit{seven equation model} implemented in the CEDRE code and the DNS results from the ARCHER code.

It showed the limits of diffuse interface models to capture complex liquid structures such as ligaments, rings or deformed droplets and encourages to add a sub-scale description of the interface dynamics through geometric variables such as the interfacial area density, the mean and Gaussian curvatures as proposed in \citeay{Cordesse_2018_NASATM_LAP}. %
\subsection*{Acknowledgment}
The support of CNES and ONERA through a PhD grant for P.Cordesse, the support of the Hadamard Doctoral School of Mathematics (EDMH), the help of L.H. Dorey (ONERA) and M. Th\'eron (CNES) are gratefully acknowledged. 

\printbibliography

\end{document}